%% file: gershon.tex
\documentclass{moriond}

\usepackage{ifthen} 
\newboolean{pdflatex}
\setboolean{pdflatex}{true} 

\newboolean{articletitles}
\setboolean{articletitles}{false} 

\newboolean{uprightparticles}
\setboolean{uprightparticles}{false} 

\newboolean{inbibliography}
\setboolean{inbibliography}{false} 

\usepackage{amsmath} 
\usepackage{amssymb}
\usepackage{amsfonts}
\usepackage{upgreek} 
\usepackage{hyperref}    
\usepackage[hypcap=true]{caption}

\include{lhcb-symbols-def}

\usepackage{cite} 
\usepackage{mciteplus}

\newcommand{\Photo}{\vspace*{3mm}\includegraphics[height=35mm]{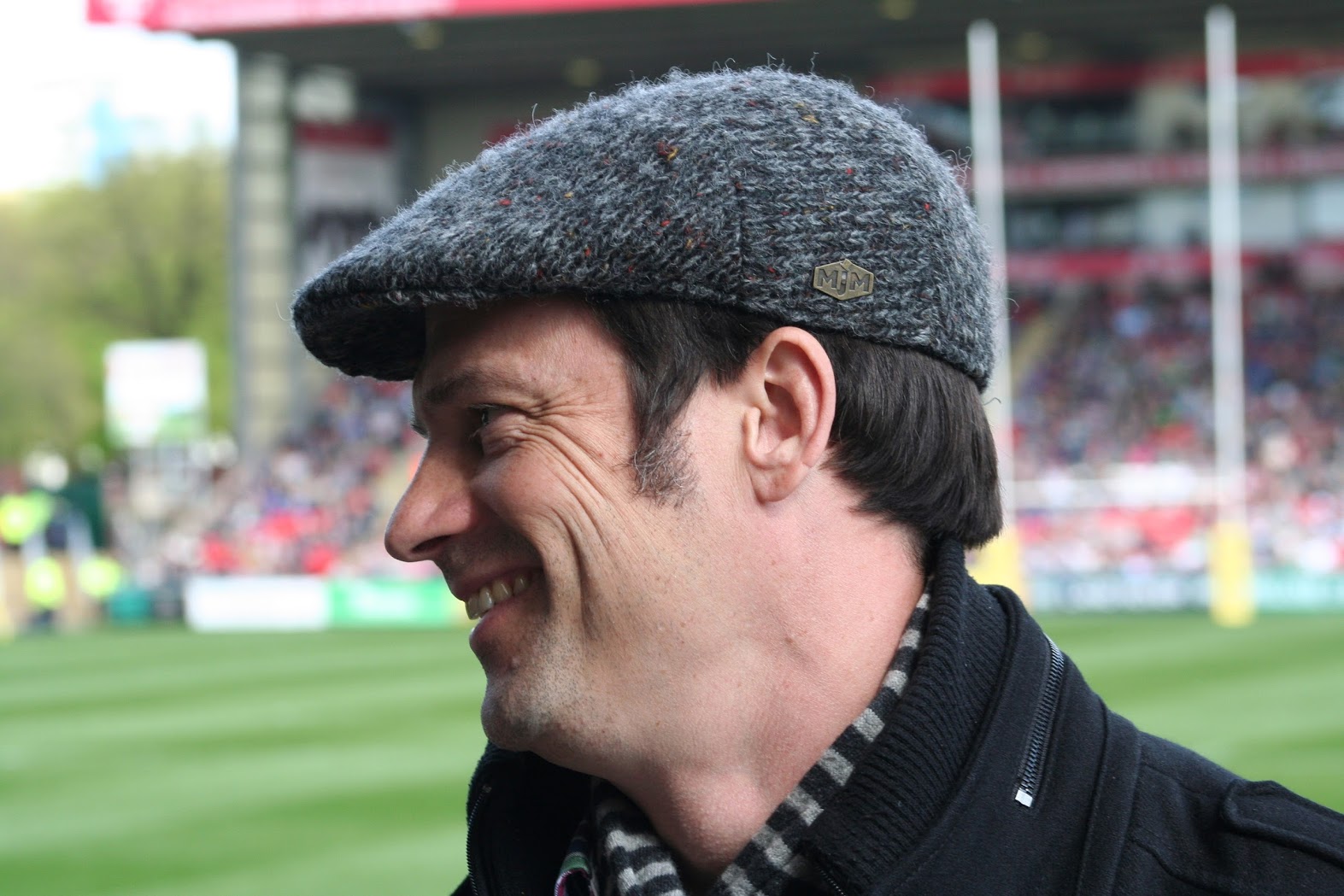}}

\begin{document}
\vspace*{4cm}
\title{Experimental summary}
\author{T.J.~Gershon}
\address{Department of Physics, University of Warwick, Coventry, United Kingdom}
\maketitle\abstracts{
A summary, from an experimental perspective, of the $52^{\rm nd}$ Rencontres de Moriond session on Electroweak Interactions and Unified Theories is presented.}

\section{Introduction}

The $52^{\rm nd}$ Rencontres de Moriond session on Electroweak Interactions and Unified Theories provided, in the usual tradition, a forum for the presentation of a wide range of new results in experimental particle physics.
In addition, vibrant discussions continued throughout the week, both in the sessions themselves, on ski lifts and pistes, and over meals, coffees and other beverages.  
In the age of social media, the discussion is no longer limited to those present at the meeting, and physicists (and sometimes other interested parties) from around the world have also had their say, for example by following {\tt @\_Moriond\_} on Twitter.
In the following, these new results and some of the resulting discussion are summarised.
For reasons of brevity, and to preserve the sanity of the author, many interesting topics are unfortunately excluded from the summary.

To start, it is perhaps helpful to recollect the situation at the end of the $51^{\rm st}$ Rencontres de Moriond, one year ago.
The experimental summary at that time~\cite{Hoecker:2016lxt}, contained discussion of a possible signature of a new state, with mass near $750 \gev$, for which there were indications in the diphoton invariant mass spectra of both ATLAS~\cite{ATLAS-CONF-2016-018} and CMS~\cite{CMS-PAS-EXO-16-018} experiments. 
Subsequent updates of these searches with larger data samples~\cite{Aaboud:2016tru,Khachatryan:2016hje} revealed distributions that are consistent with the background-only expectation; hence the excitement surrounding this possible new state rapidly dissipated.
Among several interesting lessons which can be learned from this episode, perhaps the most important is the salience of the $5\sigma$ threshold usually required to claim an observation in particle physics.  
This is worth remembering at all times, and particularly when looking at a summary, such as this one, which is likely to focus attention on whichever anomalies happen to be causing excitement at a given time.  

\begin{figure}[!htb]
  \centering
  \includegraphics[width=0.45\linewidth]{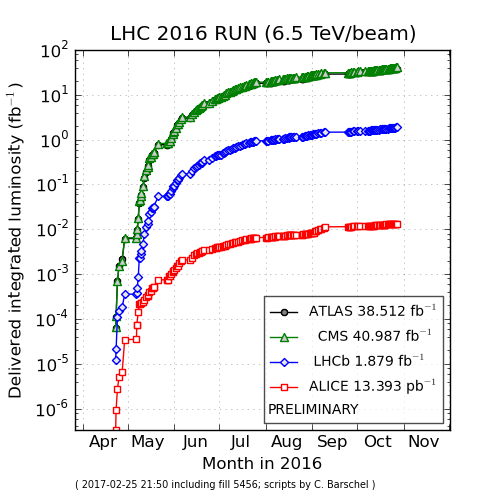}
  \includegraphics[width=0.45\linewidth,bb=0 -20 500 500,clip=true]{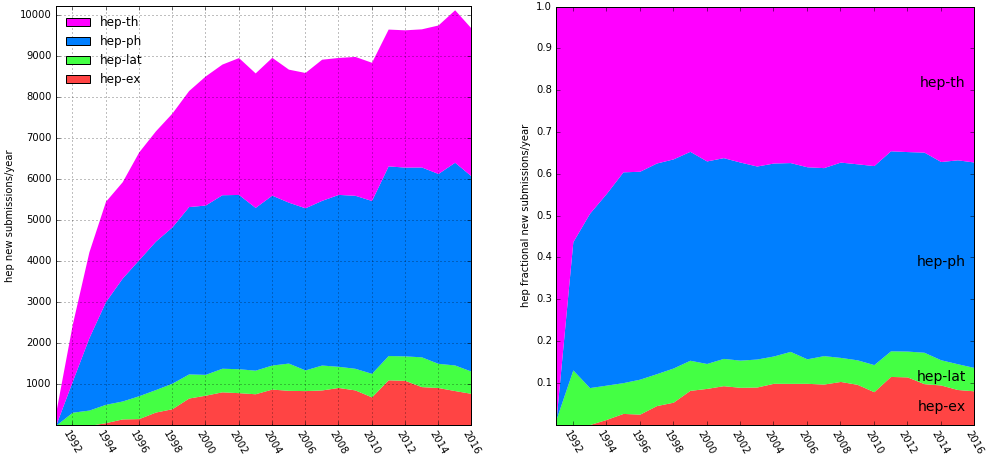}
  \caption[]{
    (Left) Delivered luminosity to the four main LHC experiments during 2016.
    (Right) Number of submissions per year in the hep-related categories of the arXiv preprint server~\cite{arXiv}.
}
  \label{fig:lumi}
\end{figure}

The absence of any clear discovery of new phenomena beyond the Standard Model (SM) at the LHC is sometimes presented as a ``failure''.  
Although this appears to be the era of fake news, our community should not allow this view to take root.  
In fact, the LHC has been astonishingly successful, with the delivered luminosity in 2016 -- shown in Fig.~\ref{fig:lumi}(left) -- so high that the experiments have needed to adjust their operations and computing models accordingly.  
The experiments in turn have also been successful, producing voluminuous publications describing their exploitation of the data.
[A downturn since 2011 in the number of submissions per year in the hep-ex category of the arXiv server~\cite{arXiv}, shown in Fig.~\ref{fig:lumi}(right), appears to be due to a decrease in the number of conference notes and proceedings that are submitted, rather than a reduction in new results.]
A (perhaps idealistic) description of an experimentalist's task is to explore what the data tell us, without bias; by any measure the LHC experiments appear to be succeeding in that task.
This exploitation of the data is indeed leading to many new discoveries, including the recent observation of five new excited $\Omega_c$ states in the same mass spectrum~\cite{Morello,LHCb-PAPER-2017-002}. 
However, it will always be possible that signals are waiting to be found and that we are simply not looking in the right place or in the right way.
New ideas to exploit better the data are always needed.
Thus, while we certainly have not failed, it is also true that we have not yet succeeded as much as we would like.   

\section{The BEH scalar boson}

In July 2017, we will celebrate the 5$^{\rm th}$ birthday of the SM scalar boson~\cite{Higgs:1964pj,Englert:1964et,Aad:2012tfa,Chatrchyan:2012xdj}.
Five years is still a very young age, and while the main features of the BEH boson such as mass and spin are established, our knowledge is developing at a rapid rate.
(For comparison, we are still learning a great deal of fundamental physics from the beauty quark, as it reaches its 40$^{\rm th}$ birthday, and the muon, which is now a venerable 80 years old.)

\begin{figure}[!htb]
  \centering
  \includegraphics[width=0.45\linewidth]{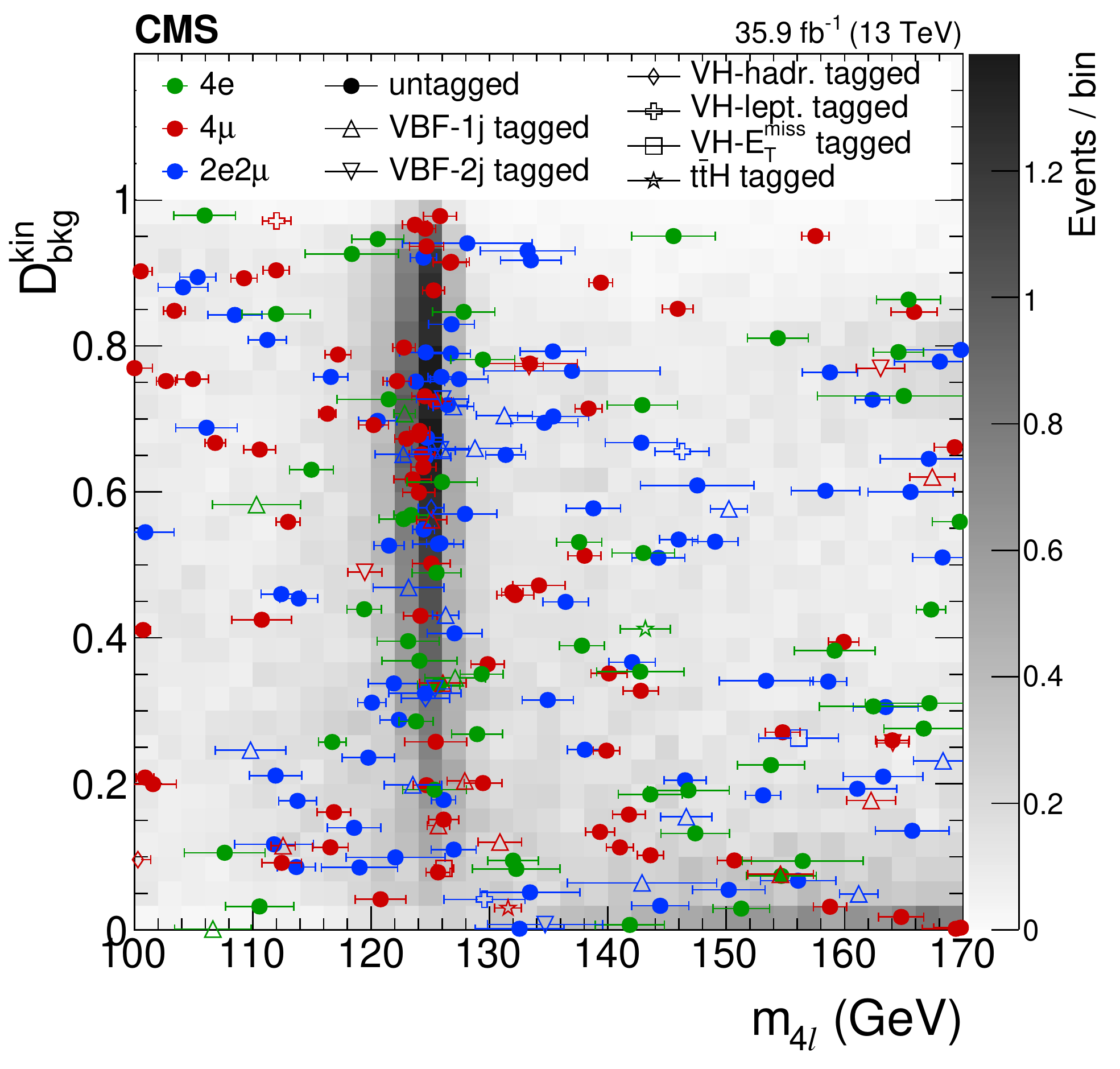}
  \includegraphics[width=0.45\linewidth]{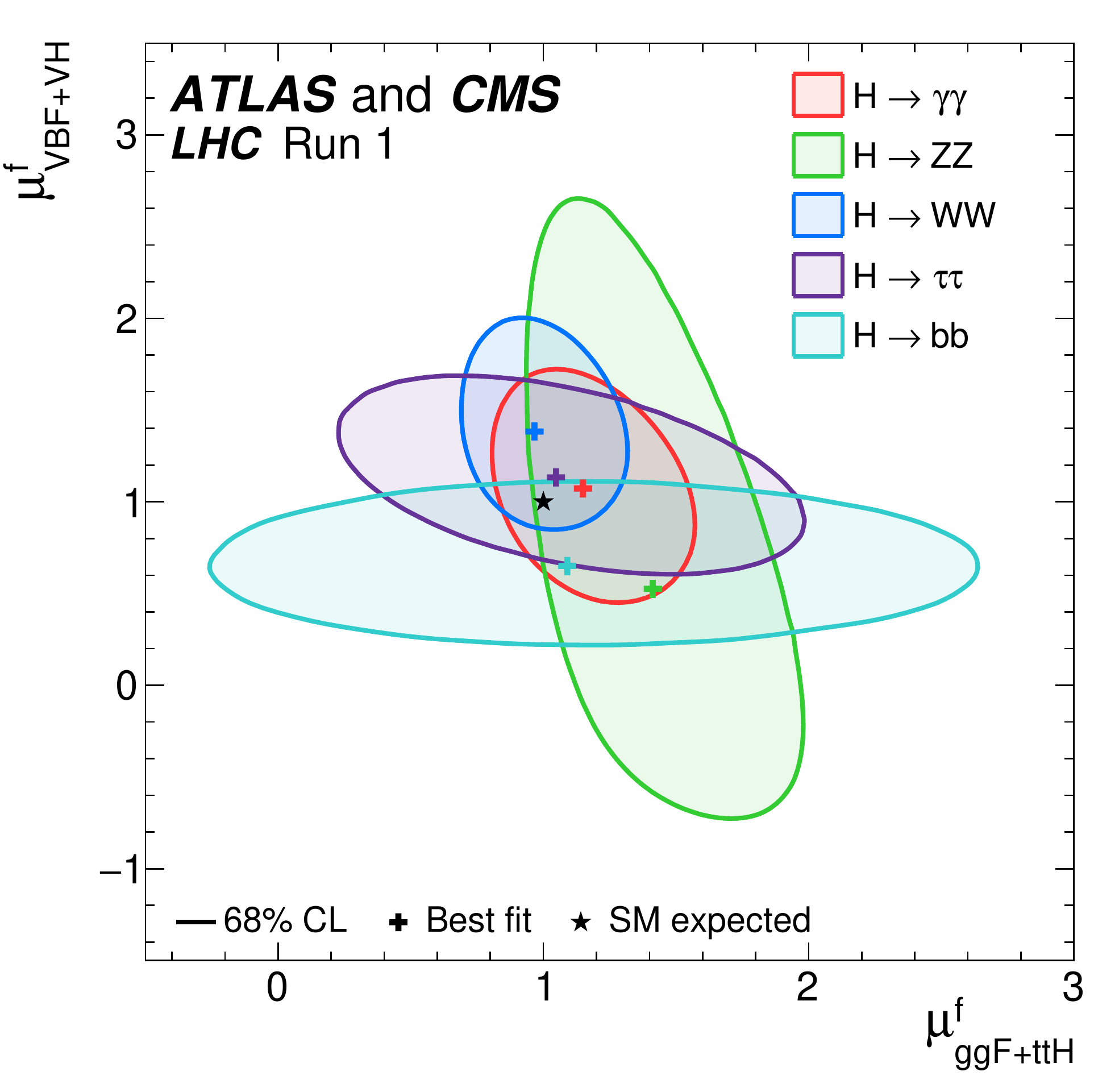}
  \caption[]{
    (Left) Selected candidates contributing to the determination of $m_H$ from $H \to ZZ^* \to 4\ell$ decays~\cite{Oda,CMS-PAS-HIG-16-041,Sirunyan:2017exp}.
    The $y$-axis shows a variable that discriminates between signal and background, the colour and marker shape indicate event category, and the error bar is the event-by-event estimate of the uncertainty on $m_{4\ell}$.
    (Right) Contours showing contraints on BEH boson production rates, normalised to the SM expectation, for ($x$-axis) fermionic production and ($y$-axis) bosonic production~\cite{Khachatryan:2016vau}.
}
  \label{fig:BEH}
\end{figure}

The $\sqrt{s} = 13 \tev$ $pp$ data sample collected by ATLAS and CMS in 2015 and 2016 allows significant improvement in understanding the BEH boson.
A study by CMS of the $H \to ZZ^* \to 4\ell$ decays, shown in Fig.~\ref{fig:BEH}(left) provides the single best measurement of the BEH mass to date, $m_H = 125.26 \pm 0.20 \stat \pm 0.08 \syst \gev$~\cite{Oda,CMS-PAS-HIG-16-041,Sirunyan:2017exp}, corresponding to about $0.2\%$ precision.
The same analysis allows a limit to be put on the width of $\Gamma_H < 1.1 \gev$ at 95\% confidence level, still far above the SM expected value. 
This limit, obtained through on-shell production, is not as constraining as results obtained from off-shell production~\cite{Aad:2015xua,Khachatryan:2016ctc} but relies less on theoretical assumptions.
Various measurements of the production and decay rates of the BEH boson are consistent with SM expectations, as summarised in Fig.~\ref{fig:BEH}(right)~\cite{Khachatryan:2016vau}; updates of these measurements with Run~2 data will provide more detailed insight.

\begin{figure}[!htb]
  \centering
  \includegraphics[width=0.35\linewidth]{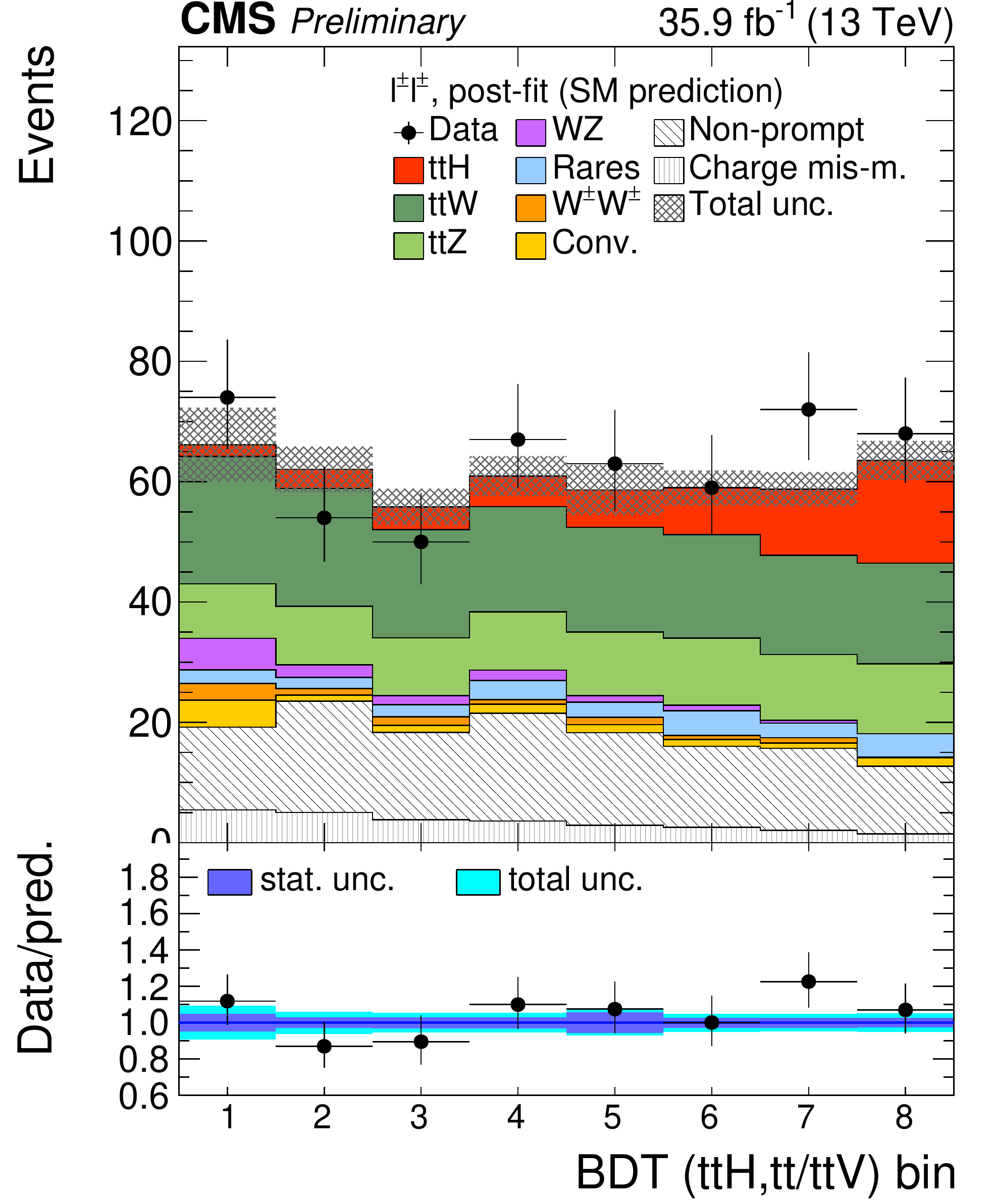}
  \includegraphics[width=0.45\linewidth]{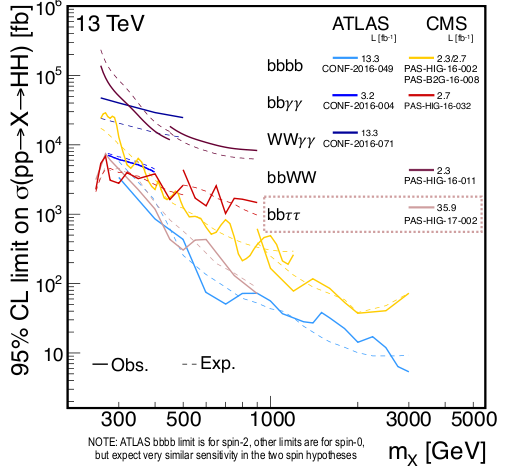}
  \caption[]{
    (Left) Combination of the multivariate classifier (BDT) outputs in the bins used for signal extraction, for the same-sign dilepton channel~\cite{CMS-PAS-HIG-17-004}.  
    Evidence for $t\bar{t}H$ production can be seen as an excess of the orange/red component at large BDT bin values.
    (Right) Limits on production of a hypothetical $X$ particle, as a function of its mass, decaying to $H\!H$, obtained through different final states~\cite{Cadamuro}.
}
  \label{fig:BEH2}
\end{figure}

It is crucial to be compare measurements and predictions of BEH couplings for bosons, quarks and leptons, and for different fermion families, but several key production and decay modes remain to be discovered.
A new result from ATLAS on $H \to \mumu$~\cite{Gaycken,ATLAS-CONF-2017-014,Aaboud:2017ojs} sets an upper limit on the cross-section times branching ratio at 2.8 times the SM prediction at the 95\% confidence level, indicating that a very interesting level of sensitivity is being reached.
There are also exciting prospects for new and improved results on the $H \to b\bar{b}, \taup\taum$ and $Z\gamma$ channels in the near future. 
Another extremely important channel is the $t\bar{t}H$ production mode, which allows the Yukawa coupling of the top quark to be determined through a tree-level process (and therefore complements existing loop-level measurements).
New results using multilepton final states have been presented by CMS~\cite{Petrucciani,CMS-PAS-HIG-17-004}; while a visible excess, corresponding to $3.3\sigma$ significance, can be seen in the same sign dilepton channel shown in Fig.~\ref{fig:BEH2}(left), combining all data it seems premature to make strong claims for signatures of $t\bar{t}H$ production.
Updates with larger Run~2 data samples are expected to allow more definitive statements.

Perhaps the most interesting as-yet-unmeasured property of the BEH boson is its self-coupling. 
Measurements of this quantity are crucial to test the shape of the BEH potential, and thus test the origin of electroweak symmetry breaking.  
A new result from CMS uses the $b\bar{b}\taup\taum$ mode, combining data in three $\taup\taum$ final states~\cite{Cadamuro,CMS-PAS-HIG-17-002,Sirunyan:2017djm}.
No significant excess over background is seen, and when interpreting the results in terms of the $H^* \to H\!H$ process an upper limit of 28 times the SM expectation is set, at 95\% confidence level.  
(The results can also be interpreted in terms of limits on the production of a hypothetical new resonance that decays to $H\!H$, as shown in Fig.~\ref{fig:BEH2}(right).)
Measurement of the Higgs self-coupling at the SM level is expected to require the full HL-LHC statistics and combination of data from ATLAS and CMS.
Determination of the quartic Higgs self-coupling will be even more challenging.

\section{Global fits to sectors of the Standard Model}
\subsection{The electroweak sector}

\begin{figure}[!htb]
  \centering
  \includegraphics[width=0.49\linewidth]{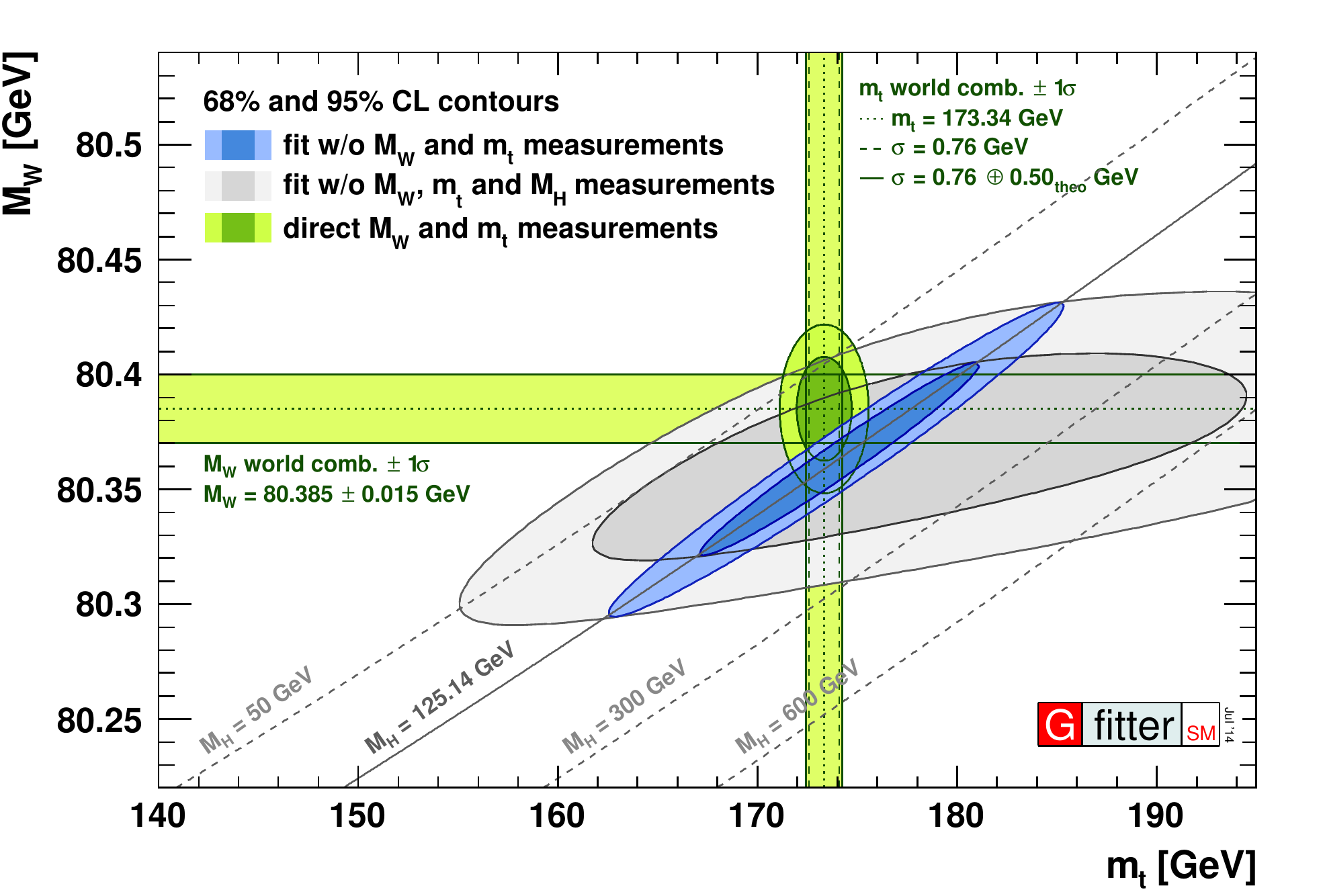}
  \includegraphics[width=0.49\linewidth]{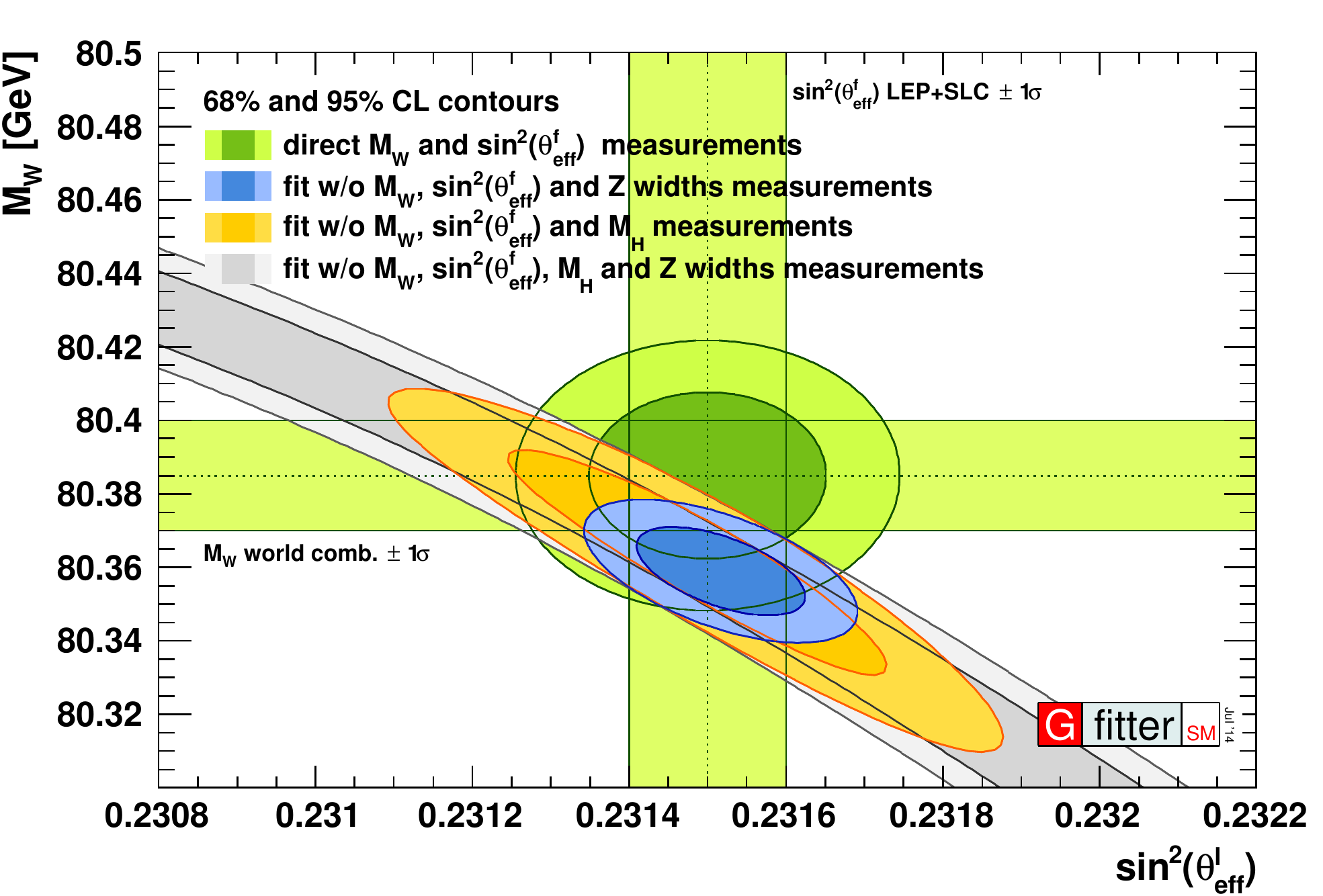}
  \caption[]{
    Results of fits to the electroweak sector of the SM~\cite{Baak:2014ora} in terms of (left) $m_W$ \vs\ $m_t$ and (right) $m_W$ \vs\ $\sin^2\left(\theta_{\rm eff}^\ell\right)$.
}
  \label{fig:EWfit}
\end{figure}

The mass of the BEH boson is one of the crucial inputs to the fit of electroweak sector of the SM~\cite{Han,Erler}, as shown in Fig.~\ref{fig:EWfit}.
The other most important inputs are:
\begin{itemize}
\item The mass of the top quark $m_t$.
  This has been measured in numerous channels by both Tevatron~\cite{Han,Bartos} and LHC~\cite{Owen} experiments.
  The most precise single measurement from the Tevatron is by the D0 experiment in the lepton+jets channel~\cite{Abazov:2015spa}; this result therefore has the largest weight in both a Tevatron combination that results in $m_t = 174.30 \pm 0.65 \gev$~\cite{TevatronElectroweakWorkingGroup:2016lid}, as shown in Fig.~\ref{fig:mt}(left), and a D0 combination~\cite{Abazov:2017ktz}.
  The most single precise measurement from the LHC is by the CMS experiment in the same lepton+jets channel~\cite{Khachatryan:2015hba}; this correspondingly has the largest weight in the LHC combination~\cite{LHCTopWG} shown in Fig.~\ref{fig:mt}(right).
  Although many of the measurements are systematics limited, many of the sources of systematic uncertainty are not correlated between the different analyses, and therefore there is still room for improving the precision of the average with more data.\footnote{
    Among the many interesting results not discussed in detail in this summary, there is a wealth of important new data on top quark production and decays, including studies of the $t\bar{t}V$ coupling (where $V$ is a gauge boson), and differential production cross-section determinations~\cite{Owen,Dobur}.
}

\begin{figure}[!htb]
  \centering
  \includegraphics[width=0.37\linewidth]{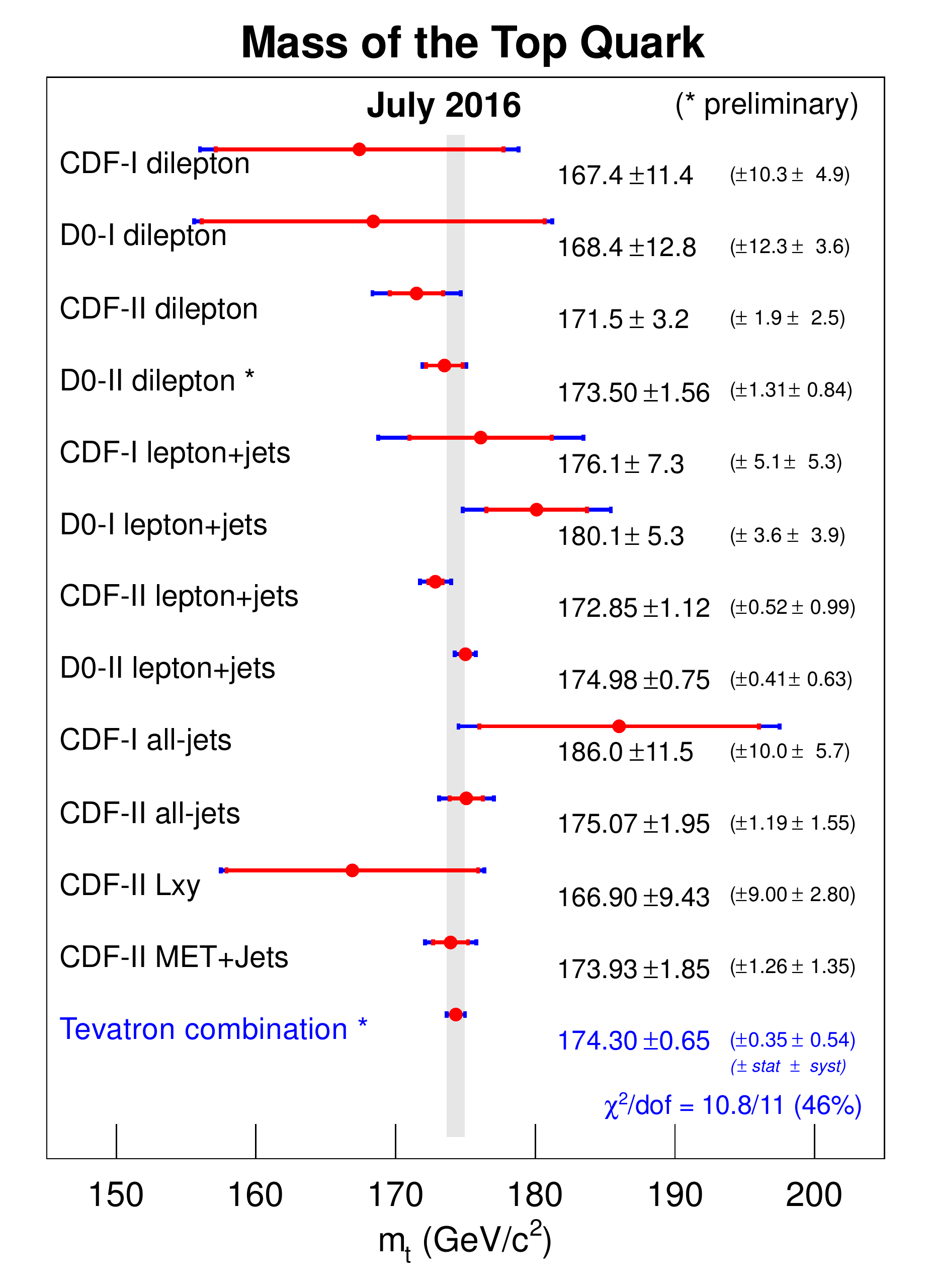}
  \includegraphics[width=0.52\linewidth]{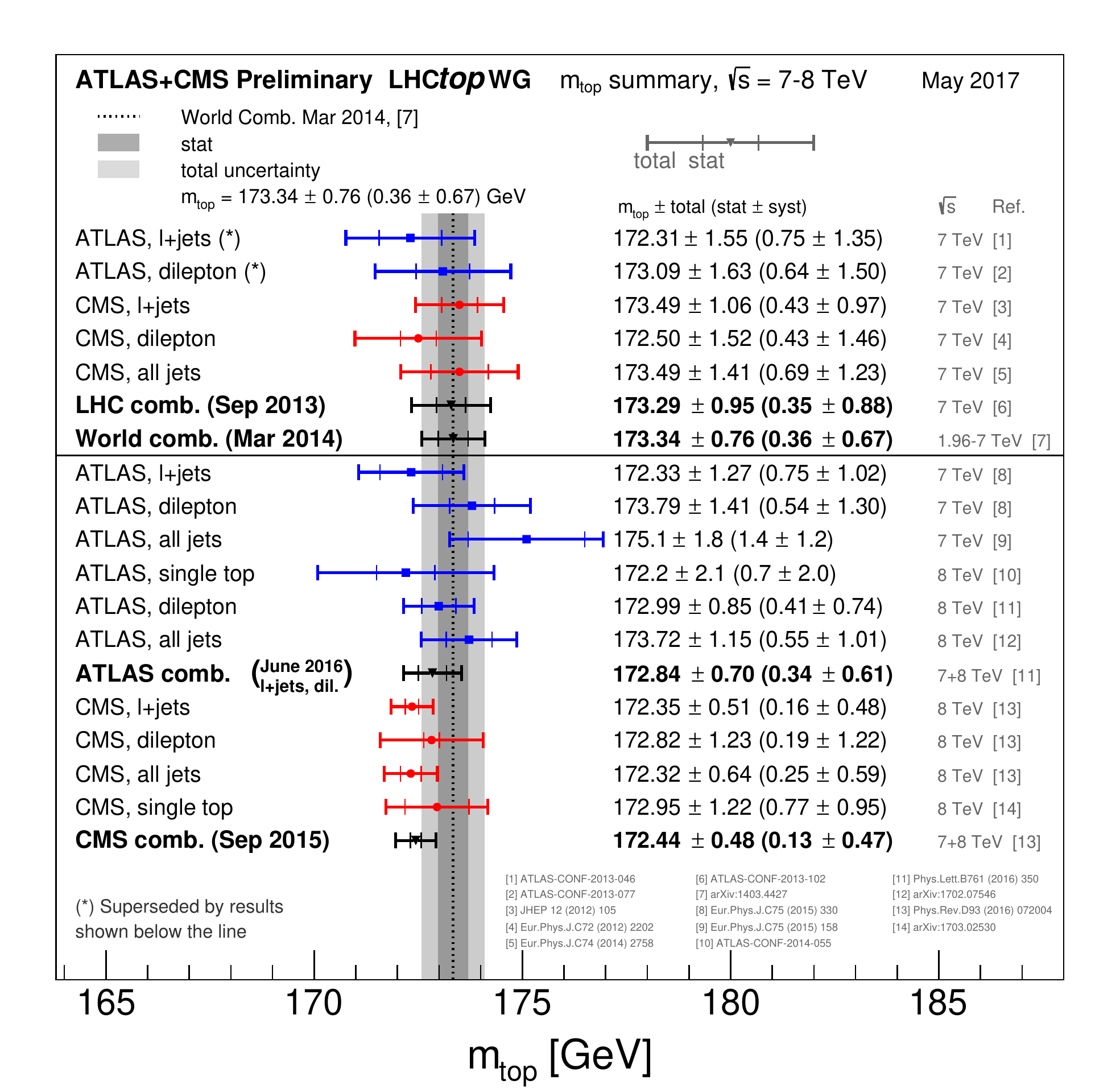}
  \caption[]{
    Combinations of determinations of the top quark mass $m_t$ from (left) the Tevatron~\cite{TevatronElectroweakWorkingGroup:2016lid} and (right) the LHC~\cite{LHCTopWG} experiments.
}
  \label{fig:mt}
\end{figure}

\item The squared sine of the effective weak mixing angle, $\sin^2\!\left(\theta_{\rm eff}^\ell\right)$.
  The most precise measurements of $\sin^2\!\left(\theta_{\rm eff}^\ell\right)$ remain those from the LEP and SLD experiments~\cite{ALEPH:2005ab}, but results from the Tevatron are reaching comparable sensitivity~\cite{tevewwg}, as shown in Fig.~\ref{fig:sin2thetaW}(left)~\cite{Han,Apyan,Erler}, with further improvement still to come.
  Results from the LHC experiments are not yet as precise, but are based on subsets of the now available data samples and therefore there is good scope for improvement.  
  At energies below the $Z$ pole, several planned experiments are expected to be able to improve the precision of tests of the running of the weak mixing angle, as shown in Fig.~\ref{fig:sin2thetaW}(right).
  
\begin{figure}[!htb]
  \centering
  \includegraphics[width=0.33\linewidth]{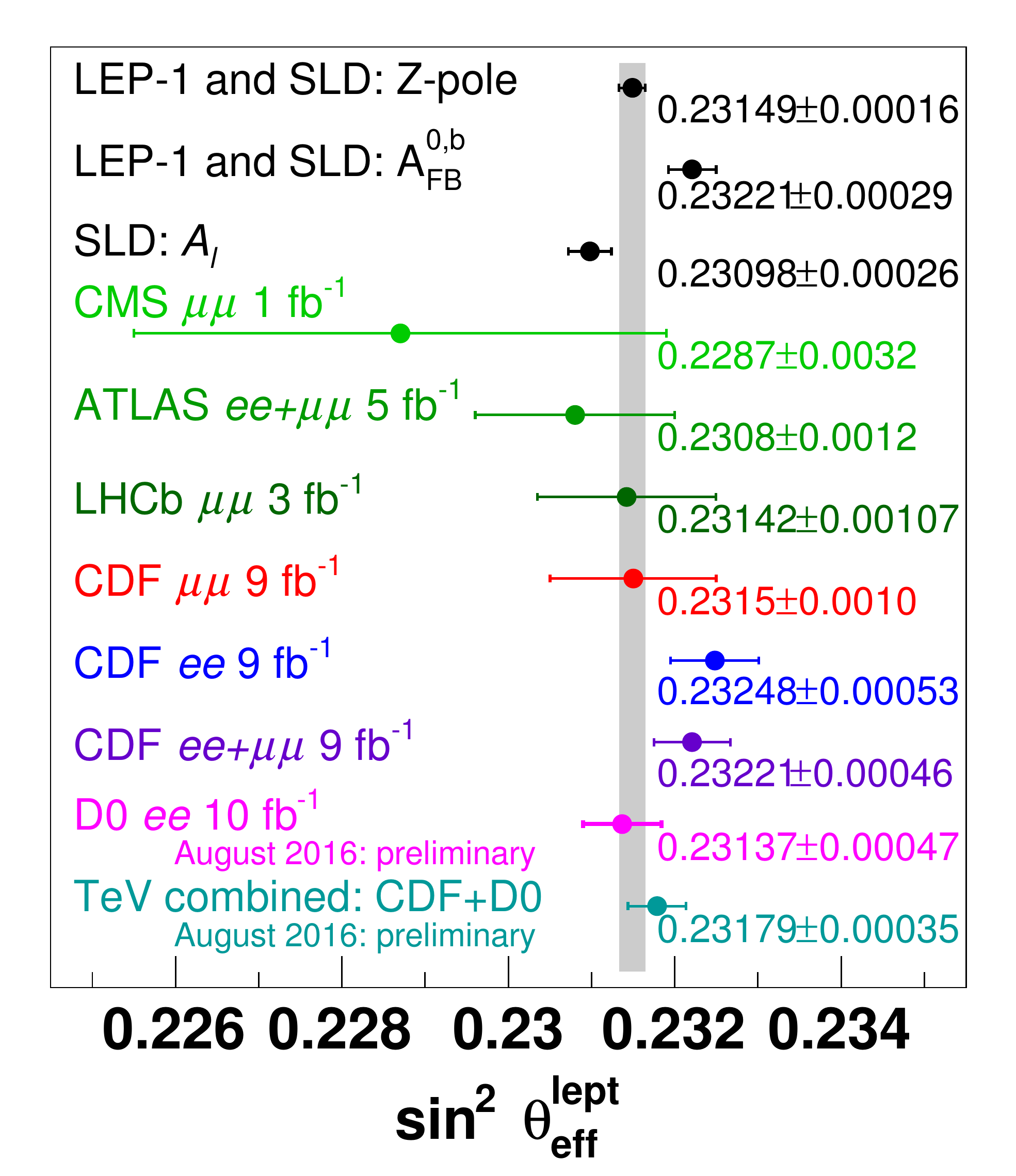}
  \includegraphics[width=0.57\linewidth]{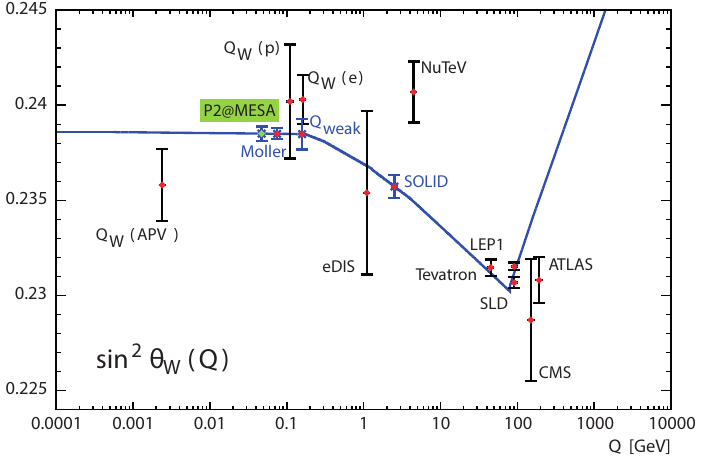}
  \caption[]{
    (Left) Comparison of experimental measurements of $\sin^2\!\left(\theta_{\rm eff}^\ell\right)$~\cite{tevewwg}.
    (Right) Current (black) and planned (blue; central value shown coinciding with the theory prediction) measurements of $\sin^2\!\left(\theta_W\right)$ at different scales $Q$~\cite{Berger:2015aaa}.
    Different measurements at the $Z$ pole are offset slightly for ease of display.
    The possible improvements in Tevatron and LHC measurements are not indicated; nor are possible determinations at $Q \approx m_{\Upsilon(4S)}$ by Belle~II.
}
  \label{fig:sin2thetaW}
\end{figure}

\item The mass of the $W$ boson, $m_W$.
  Until recently, knowledge of this quantity has been dominated by results from the Tevatron experiments, where the latest combination gives $m_W = 80.387 \pm 0.016 \gev$~\cite{Han,Aaltonen:2013iut}.
  However, the ATLAS collaboration has recently reported a result of $m_W = 80.370 \pm 0.007 \stat \pm 0.011 \mathrm{\,(exp\ syst)}\xspace \pm 0.014 \mathrm{\,(model\ syst)}\xspace \gev$ based on their data sample recorded in 2011 at a $pp$ centre-of-mass energy of $7 \tev$~\cite{Andari,Aaboud:2017svj}.
  The measurement is performed by comparing templates obtained at different values of $m_W$ to the spectra of the charged lepton transverse momentum and of the $W$ boson transverse mass in the electron and muon decay channels, as shown in Fig.~\ref{fig:mW}.
  Careful calibration of the detector response is required to minimise the experimental systematic uncertainty.
  The dominant uncertainty arises due to modelling effects, in particular due to knowledge of the parton distribution functions (PDFs) and of the $W$ \pt\ distribution~\cite{Andari,Rolandi}.
  These uncertainties can be reduced in future, although significant effort including special LHC runs will be necessary.
  Measurements in different fiducial regions will also be important~\cite{Bozzi:2015zja}.\footnote{
    There are also a huge number of results related to electroweak boson production that are not mentioned in this summary. 
    Among the most interesting are measurements of the differential production rates, with observations of $VV$ production with $13 \tev$ data expected soon~\cite{Duric,Aaboud:2017fye,Aaboud:2017pds}.
}

\begin{figure}[!htb]
  \centering
  \includegraphics[width=0.45\linewidth]{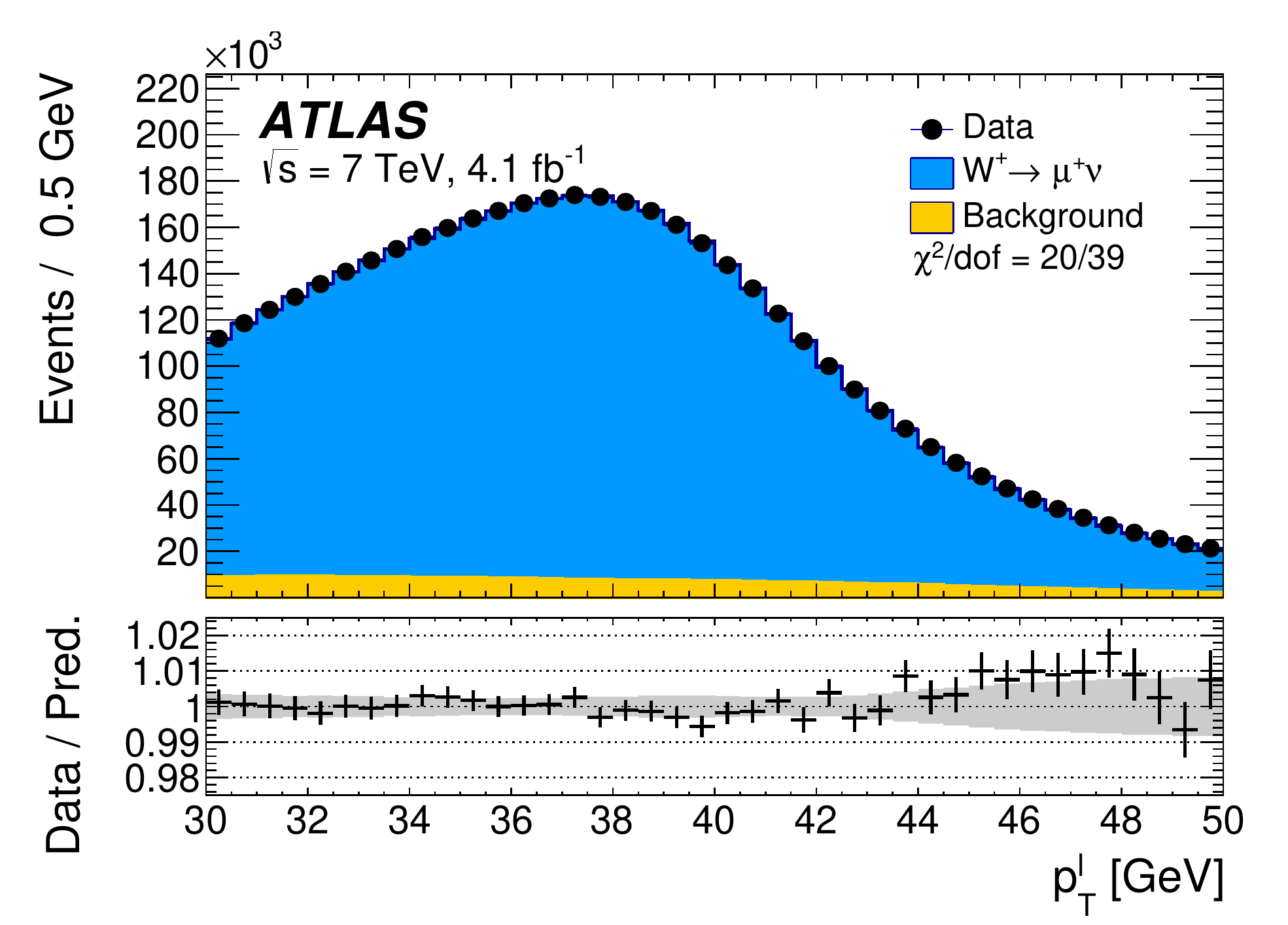}
  \includegraphics[width=0.45\linewidth]{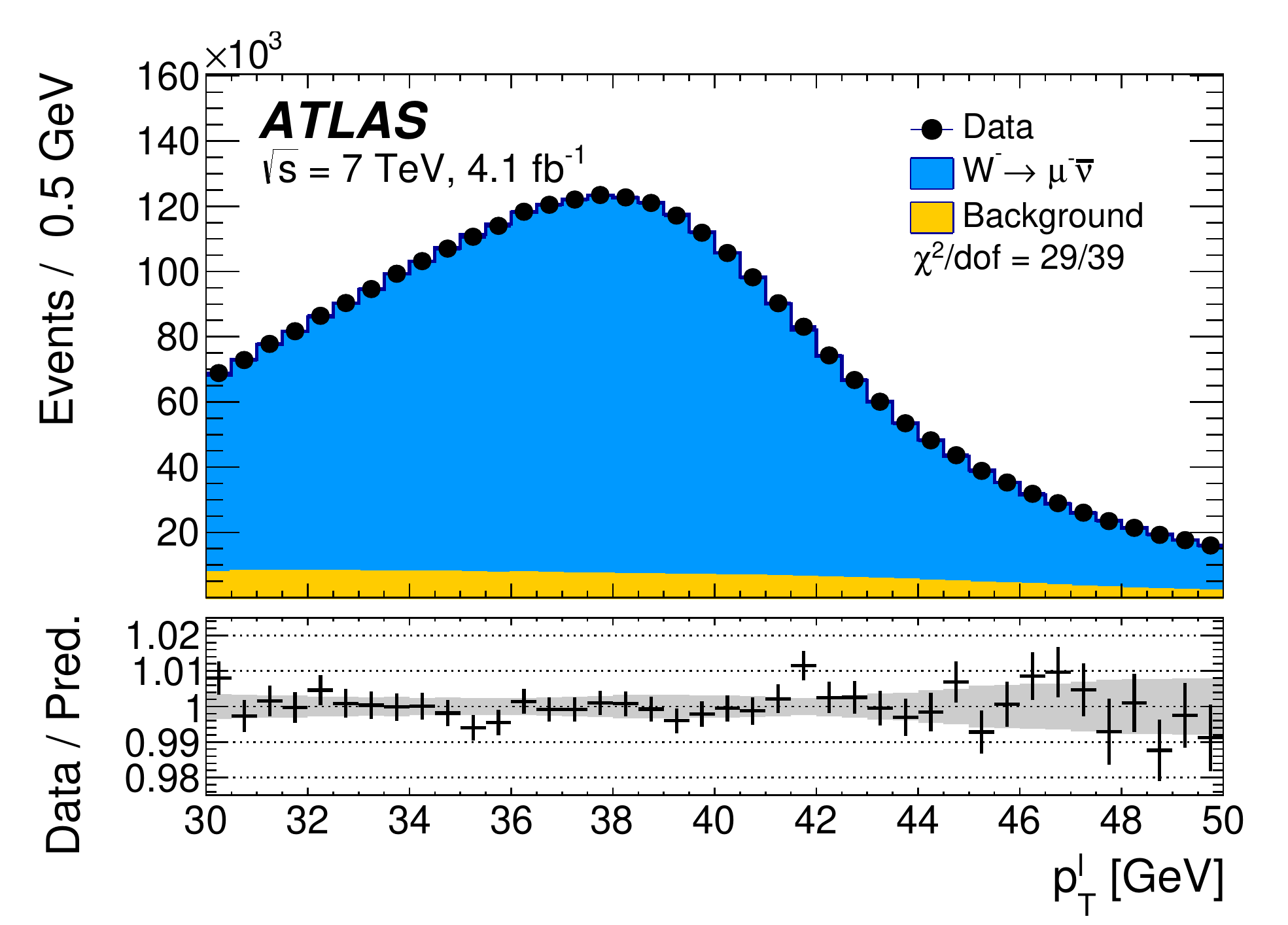} \\
  \includegraphics[width=0.45\linewidth]{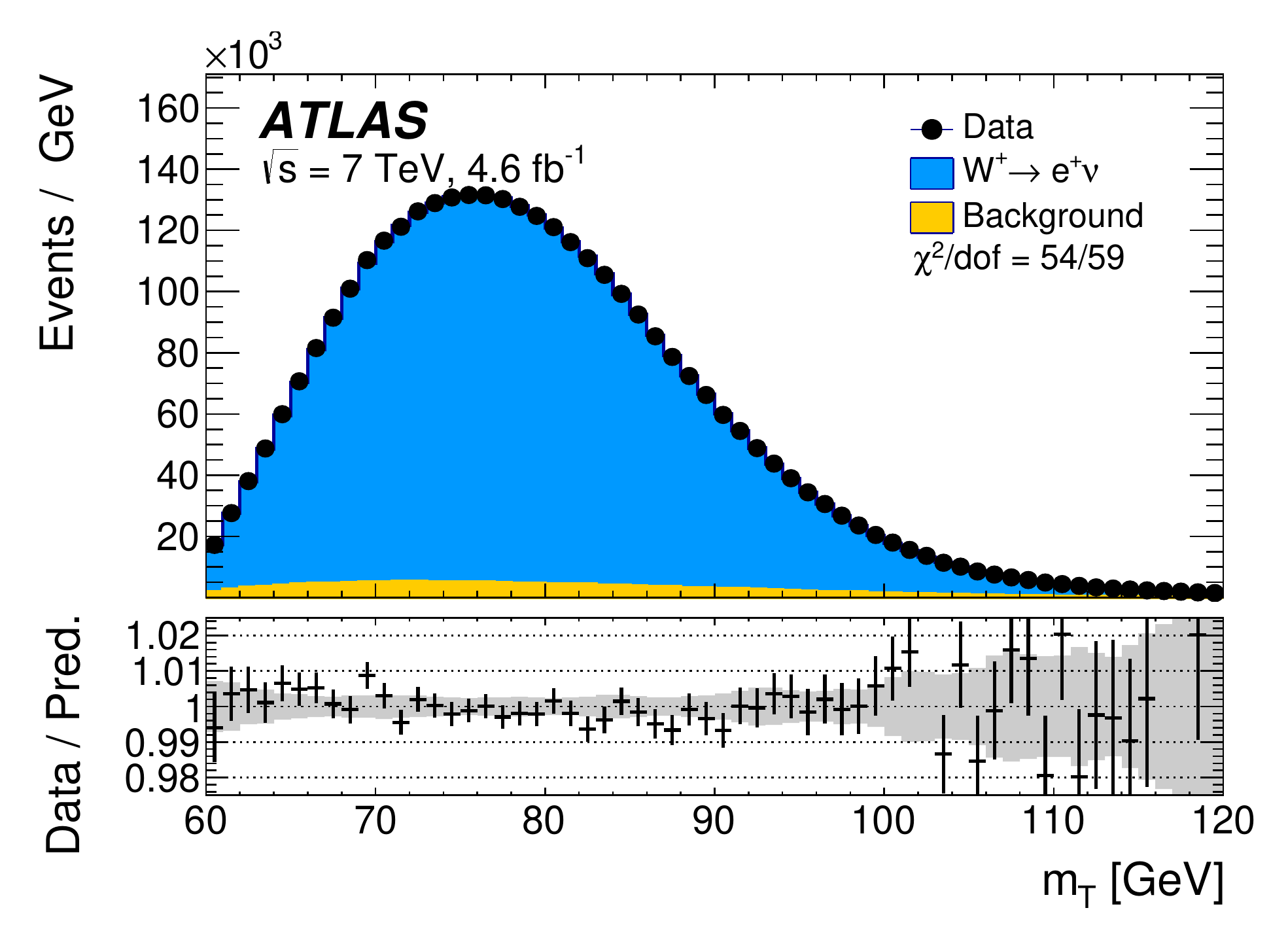}
  \includegraphics[width=0.45\linewidth]{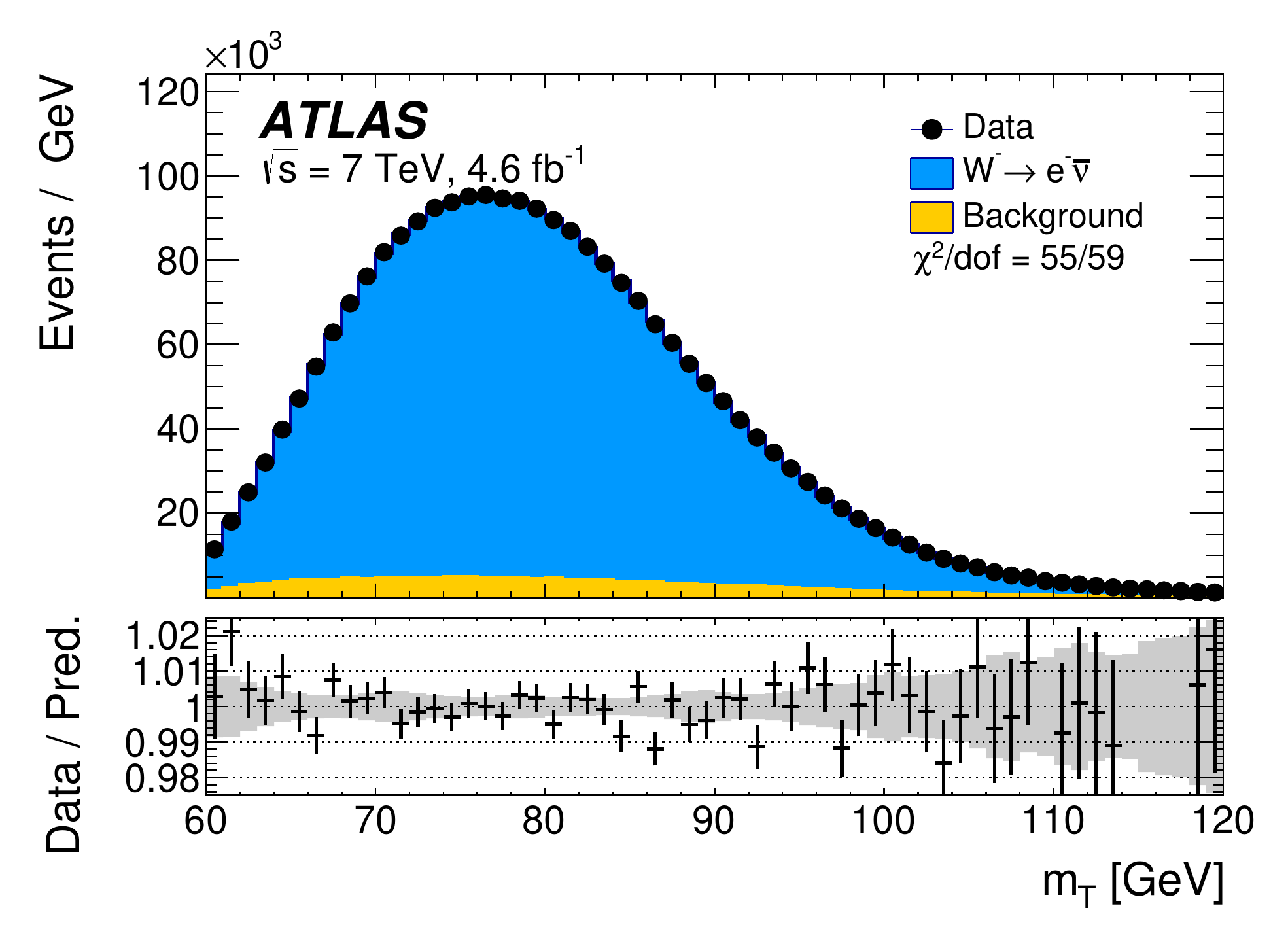}
  \caption[]{
    (Top) $\pt$ distributions for candidate muons and (bottom) $m_{\rm T}$ distribution for candidate electrons from $W$ decay~\cite{Aaboud:2017svj}.
    The figures are separated into the cases with (left) positively and (right) negatively charged leptons.  
}
  \label{fig:mW}
\end{figure}
  
\end{itemize}

The issue of PDF uncertainties that affects the $m_W$ measurement is one that has impact across a broad range of LHC measurements.  
As the precision improves various subtle effects, such as the impact of the charm mass, will need to be considered carefully.  
Nonetheless, numerous observables can be used to help reduce such uncertainties, and there are therefore good prospects for data-driven progress~\cite{Apyan,Forte,Rolandi}.
Another interesting example of LHC data being used to constrain similar types of uncertainties comes from an LHCb measurement of $\bar{p}$ production in $p{\rm He}$ collisions~\cite{Graziani,LHCb-CONF-2017-002}.
This is achieved by injecting a small amount of gas into the interaction region, using a system designed to allow measurements that help to better understand the LHC beam profile~\cite{FerroLuzzi:2005em,LHCb-PAPER-2014-047}.
The results will help to constrain uncertainties the limit the interpretation of the $\bar{p}/p$ ratio in cosmic rays~\cite{Adriani:2012paa,Aguilar:2016kjl}.
One more example of a related measurement is that of the proton-air cross-section, which has been determined at high energies by the Pierre Auger observatory~\cite{Colalillo,Collaboration:2012wt,Ulrich:2015yoo}.

\subsection{The CKM matrix}

\begin{figure}[!htb]
  \centering
  \includegraphics[width=0.42\linewidth]{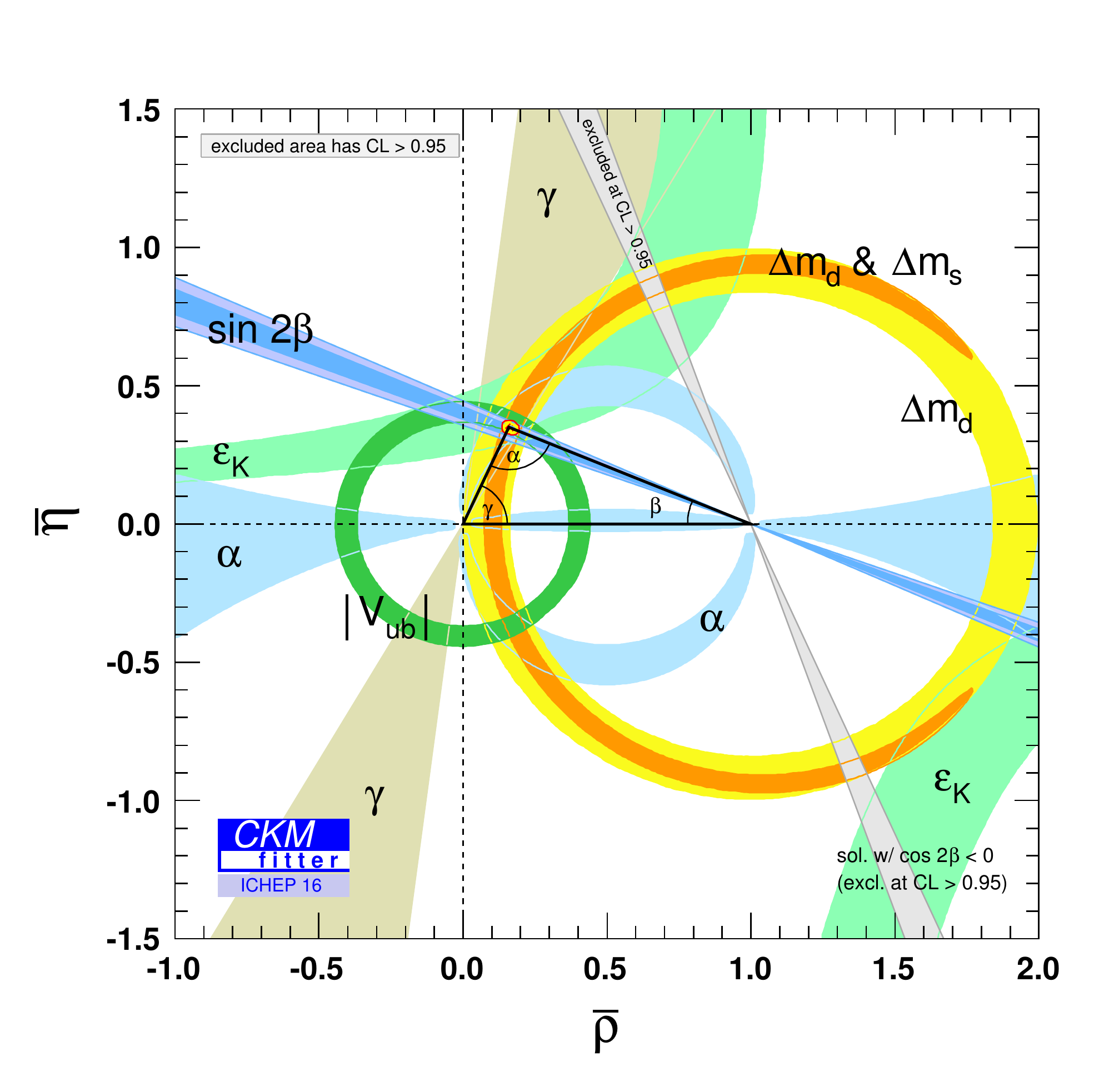}
  \includegraphics[width=0.57\linewidth]{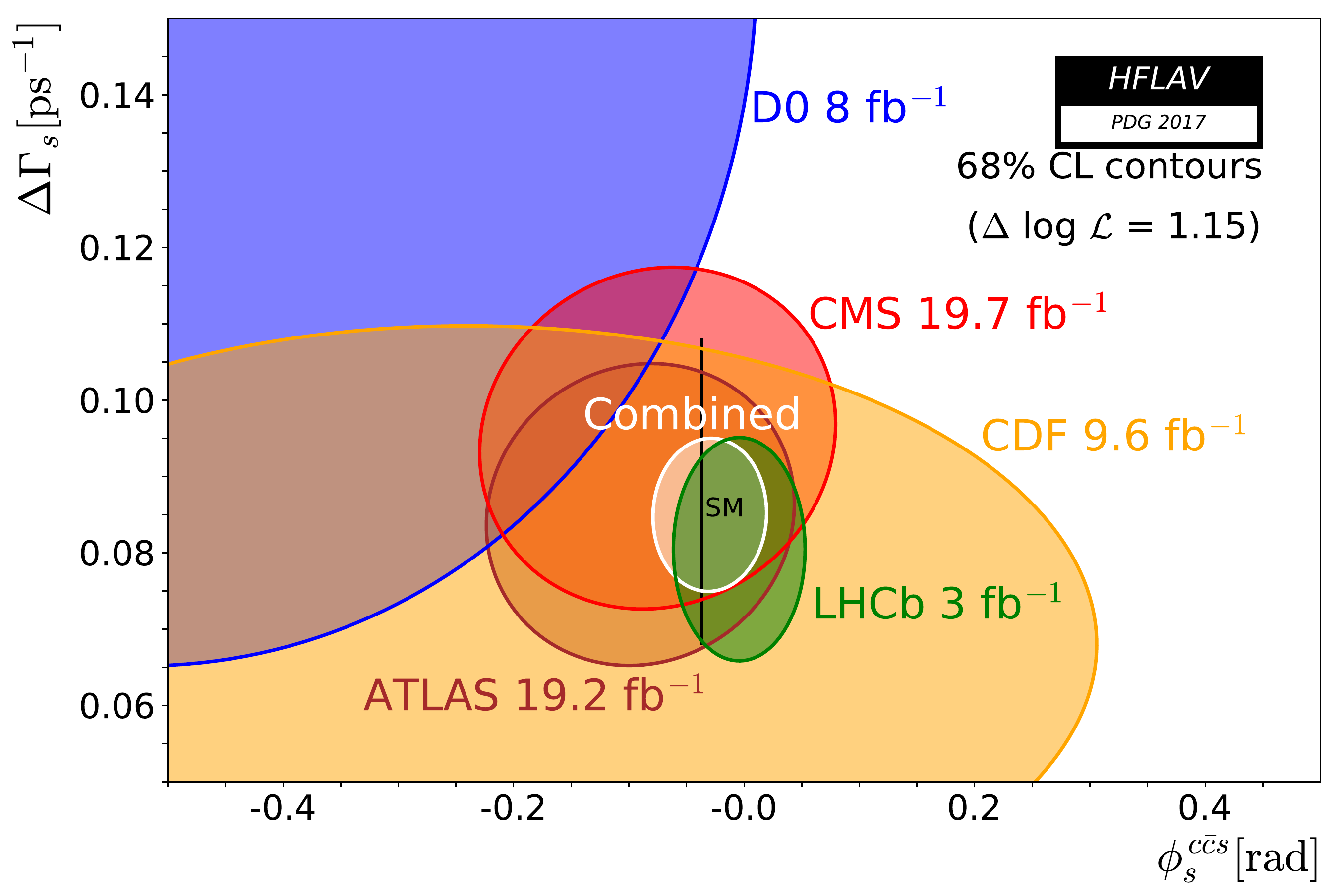}
  \caption[]{
    (Left) Constraints on the apex of the Unitarity Triangle $(\bar{\rho},\bar{\eta})$~\cite{Charles:2004jd}.  
    (Right) World average of determinations of $\phi_s$ in $b \to c\bar{c}s$ traansitions such as $\Bs \to \jpsi\phi$, presented in the plane with the width difference in the \Bs\ meson system, $\Delta \Gamma_s$~\cite{HFLAV}.
}
  \label{fig:CKM}
\end{figure}
  
Just as the electroweak sector of the SM can be overconstrained in fits, so can the flavour sector: this is usually represented in terms of the parameters of the CKM~\cite{Cabibbo:1963yz,Kobayashi:1973fv} Unitarity Triangle shown in Fig.~\ref{fig:CKM}(left).
While both sectors are ideal for precision tests, in the flavour sector most measurements are statistically limited and therefore significant further improvement can be anticipated.
Indeed, the improvement in sensitivity to new physics energy scale from the current situation to that which can be expected with results from the Belle~II experiment~\cite{Aushev:2010bq,Abe:2010gxa} and the LHCb phase 2 upgrade~\cite{CERN-LHCC-2017-003} corresponds to a factor of 2.7--4, and as such is comparable to going from an LHC $pp$ collision energy of $8 \tev$ to 21--$32 \tev$~\cite{Zupan}.

New results relevant to the CKM fit include a theoretically pristine determination of the angle $\beta$ of the Unitarity Triangle using $B \to Dh^0$ and subsequent $D \to \KS\pip\pim$ decays, where the known strong phase variation across the Dalitz plot of the neutral $D$ meson decay allows both $\sin(2\beta)$ and $\cos(2\beta)$ to be measured~\cite{Bondar:2005gk}.
The new measurement is obtained from a simultaneous joint analysis of BaBar and Belle data, and allows the ambiguous solution for $\sin(2\beta)$ with $\cos(2\beta)<0$ to be definitively ruled out~\cite{Rohrken}.
The angle $\gamma$ is now known to around $6^\circ$ precision, thanks to new results on $B \to DK$ and related modes from LHCb~\cite{Carson,LHCb-PAPER-2016-032,HFLAV}, and a new result on the weak phase difference between $\Bs \to \jpsi\Kp\Km$ decays with and without oscillations, obtained from the high $\Kp\Km$ mass region~\cite{LHCb-PAPER-2017-008}, improves the world average as shown in Fig.\ref{fig:CKM}(right).

In the charm sector, constraints on \CP\ violation in $\Dz$--$\Dzb$ oscillations are becoming increasingly precise, mainly due to new results from LHCb~\cite{Morello,LHCb-PAPER-2016-063,Weidenkaff}; however the interpretation of these constraints is limited by a lack of knowledge of the mixing parameter $x_D = \Delta m_D / \Gamma_D$, which is consistent with zero within about $2\sigma$~\cite{HFLAV}.
(Here $\Delta m_D$ is the mass difference between the physical eigenstates and $\Gamma_D$ is their average width.)
It is of utmost importance for the charm physics community to improve the knowledge of $x_D$ so that future prospects for \CP\ violation searches can be reliably assessed.  

\subsection{The PMNS matrix}

\begin{figure}[!htb]
  \centering
  \includegraphics[width=0.45\linewidth]{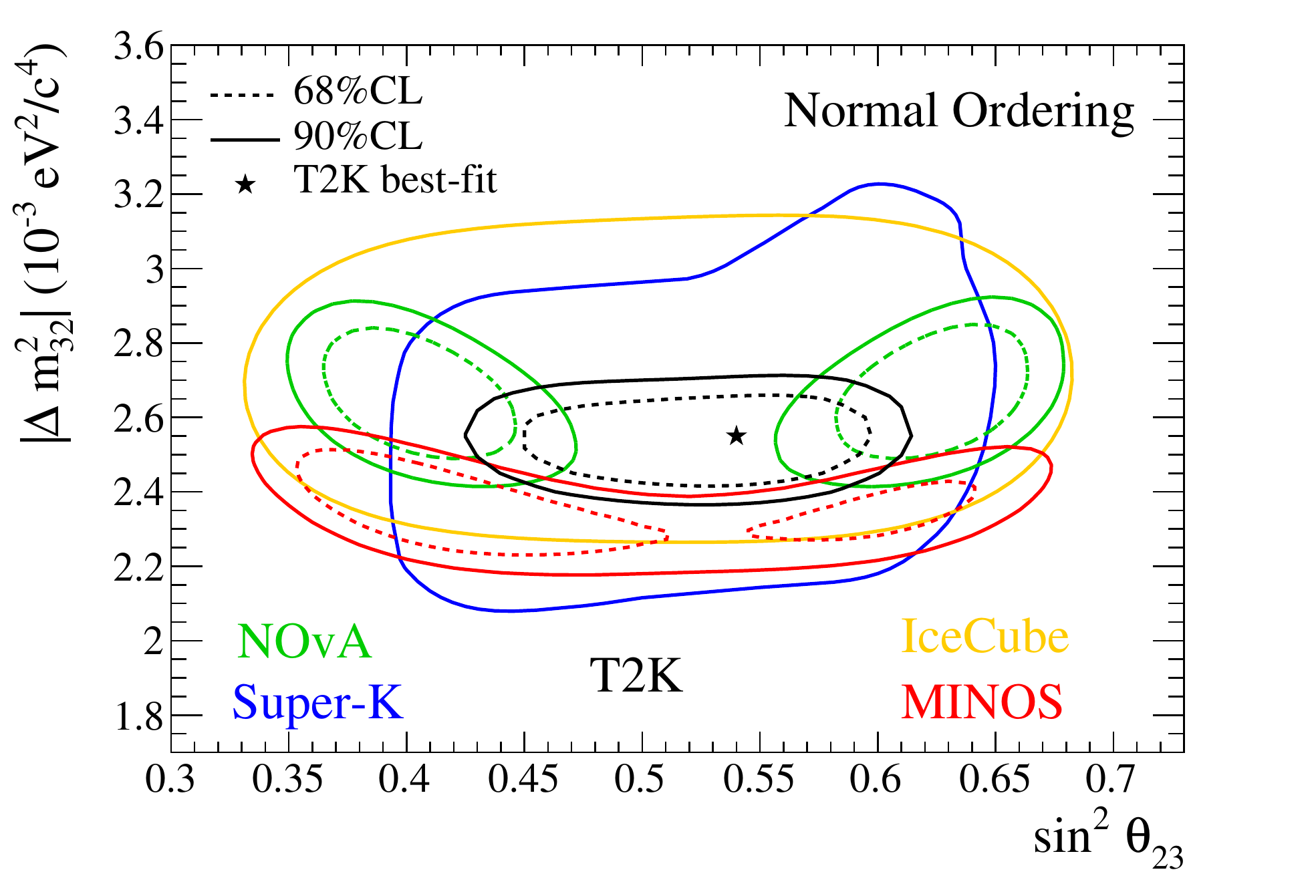}
  \includegraphics[width=0.45\linewidth]{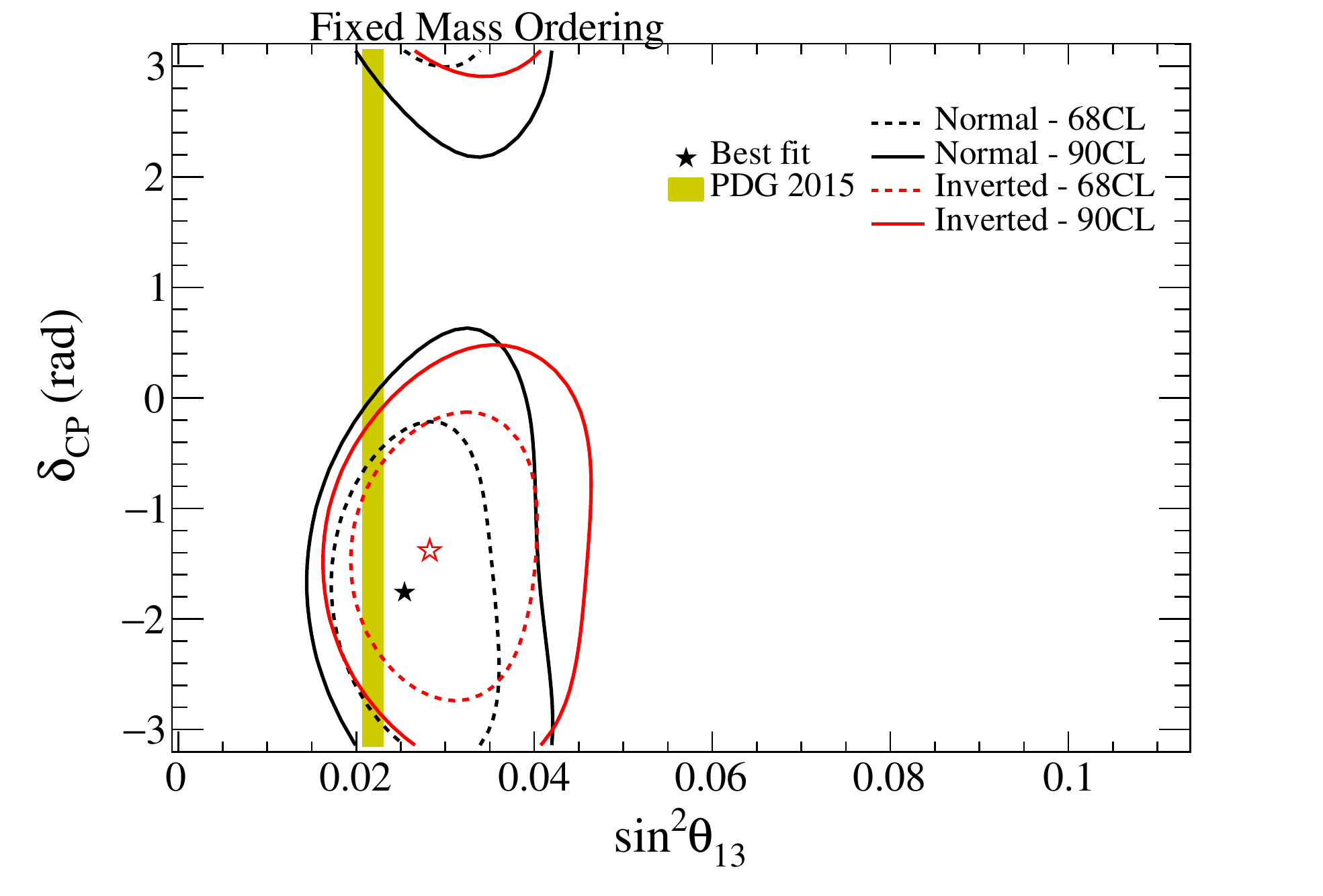}
  \caption[]{
    (Left) Constraints in the plane of $\Delta m^2_{32}$ \vs\ $\sin^2\!\left(\theta_{23}\right)$ from T2K and other experiments~\cite{Abe:2017vif}.
    (Right) Constraints in the plane of $\delta_{\CP}$ \vs\ $\sin^2\!\left(\theta_{13}\right)$ from T2K~\cite{Abe:2017vif}, compared to the world average for the latter~\cite{PDG2014,An:2015rpe}.
}
  \label{fig:PMNS}
\end{figure}

Mixing in the lepton sector is described by the PMNS~\cite{Pontecorvo:1957qd,Maki:1962mu} matrix, in close analogy to the CKM matrix for the quark sector.
The analogy is not perfect, since neutrinos are massless in the SM, but holds for the current discussion.
Measurements of the parameters of the PMNS matrix are high priorities in neutrino physics, with particular focus on the \CP\ violating phase $\delta_{\CP}$.
The latest results from the long-baseline oscillation experiments with off-axis detectors No$\nu$A and T2K, as well as from the reactor experiments Daya Bay and Double Chooz~\cite{Jediny,Nakadaira,Meregaglia,Carroll}, give precision on $\sin^2\!\left(\theta_{13}\right)$ that makes it now the best measured neutrino mixing angle.
The octant of $\theta_{23}$, \ie\ whether $\sin^2\!\left(\theta_{23}\right)$ is $>$ or $<0.5$ is still to be determined, as shown in Fig.~\ref{fig:PMNS}.  
As regards the mass hierarchy, normal ordering is preferred over inverted ordering, but not yet at a definitive level of significance~\cite{Schwetz}.
Similarly, while current data disfavour certain values of $\delta_{\CP}$, global analyses do not give a significant rejection of \CP\ conservation.  
More data and upgrades at No$\nu$A and T2K, and the next generation experiments DUNE and HyperK, will give significant improvements in sensitivity and should be able to resolve all these issues.

\begin{figure}[!htb]
  \centering
  \includegraphics[width=0.46\linewidth]{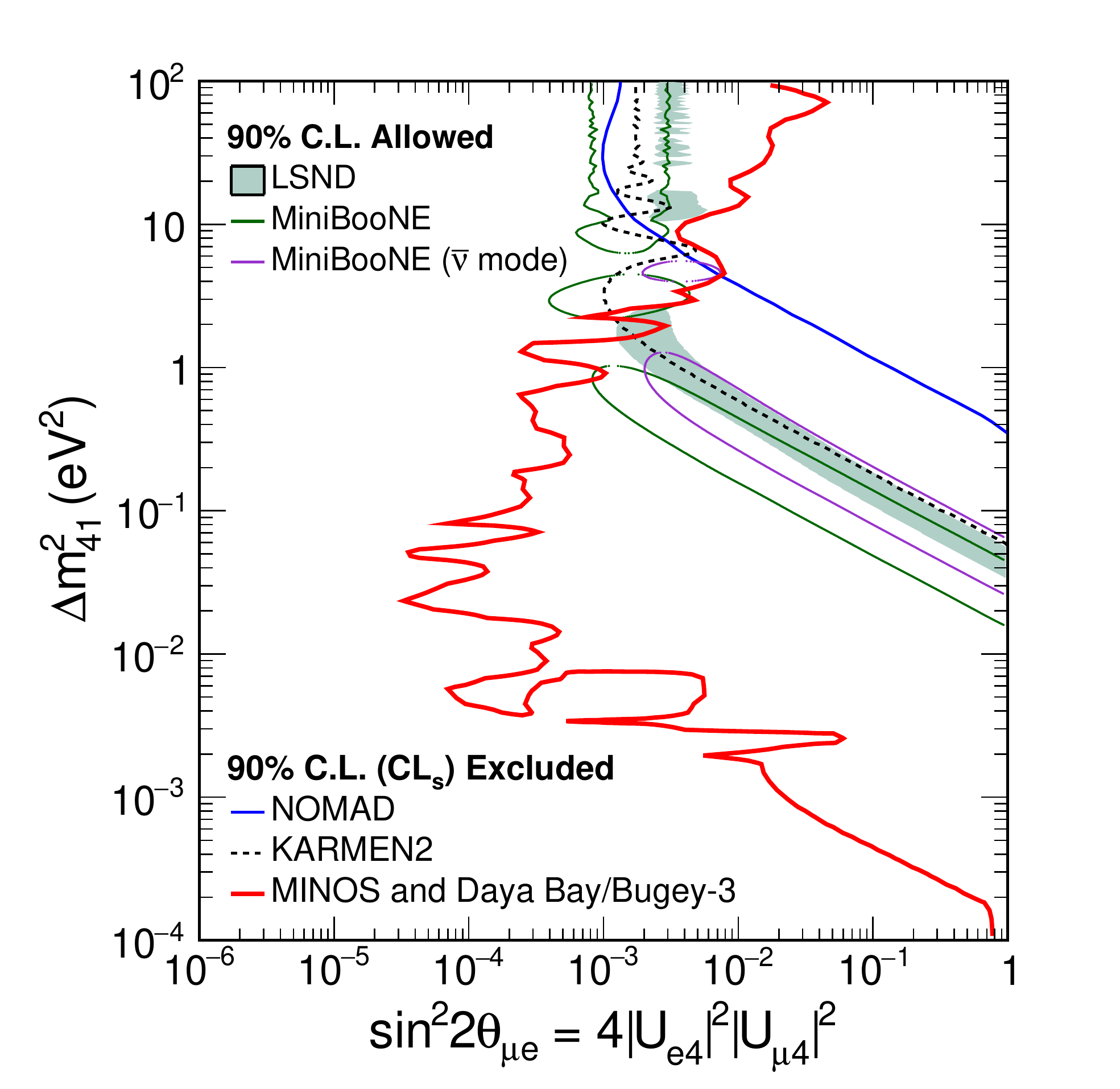}
  \includegraphics[width=0.43\linewidth]{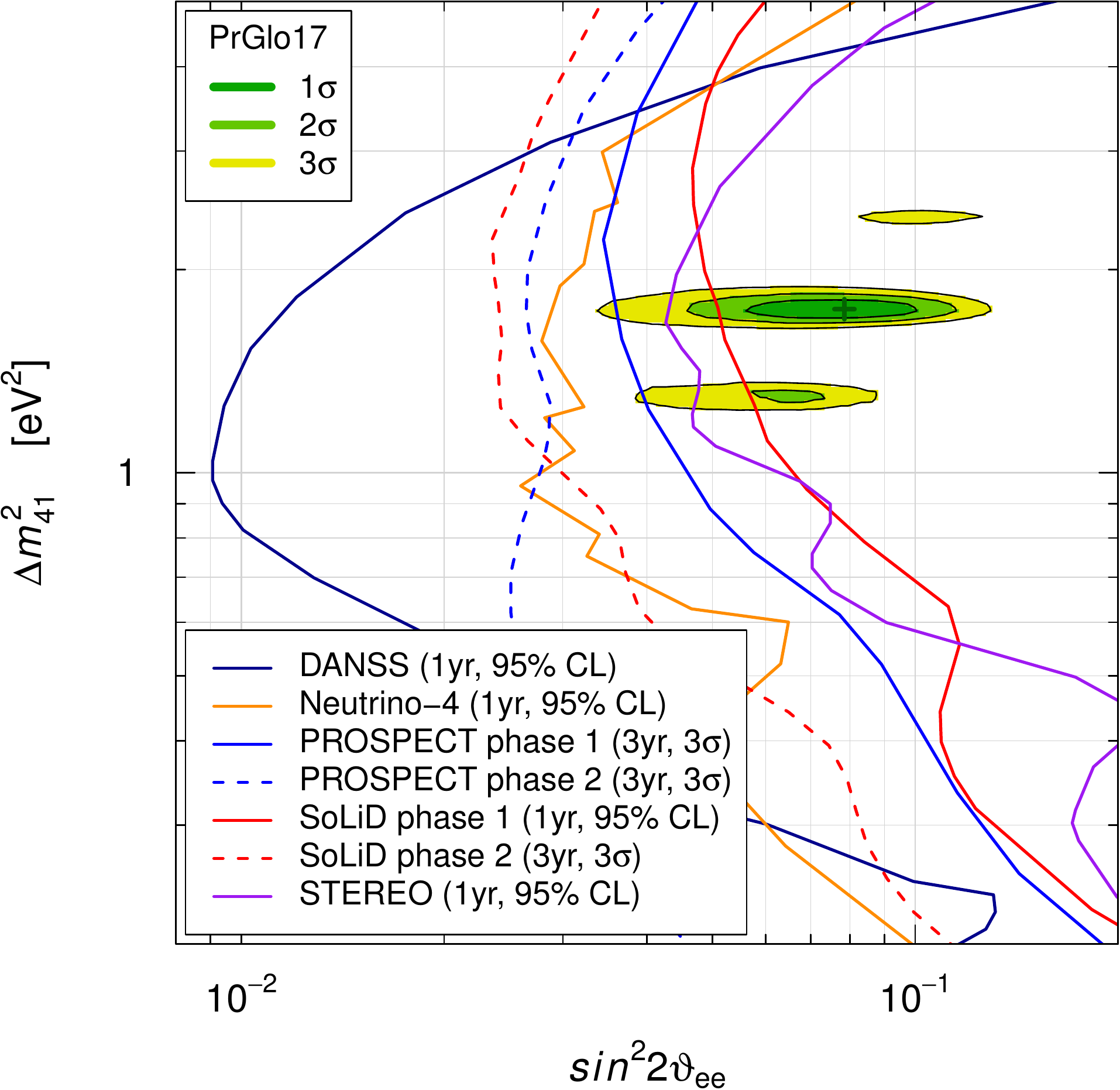}
  \caption[]{
    (Left) Constraints in the plane of $\Delta m^2_{41}$ \vs\ $\sin^2\!\left(2\theta_{\mu e}\right)$ obtained from combined analysis of MINOS, Daya Bay and Bugey-3 data~\cite{Adamson:2016jku}, compared to limits and allowed regions from previous experiments and global fits.
    (Right) Sensitivity in the $\Delta m^2_{41}$ \vs\ $\sin^2\!\left(2\theta_{ee}\right)$ plane from several very short baseline reactor neutrino experiments compared to regions preferred by global fits~\cite{Gariazzo:2017fdh}.  
}
  \label{fig:stNu}
\end{figure}

The existence of sterile neutrinos would cause the PMNS matrix to be non-unitary, as the $e$, $\mu$ and $\tau$ neutrinos would mix with the extra state(s). 
Several results have hinted at such anomalous mixing, motivating a number of new experiments to provide improved measurements~\cite{Carroll,Danilov,Giunti,PedrodeAndre}.  
One particularly innovative analysis is the combination of data from MINOS, which is sensitive to $\sin^2\!\left(\theta_{24}\right)$~\cite{MINOS:2016viw}, and from Daya Bay, which is sensitive to $\sin^2\!\left(2\theta_{14}\right)$~\cite{An:2016luf}.
The combination allows constraints to be placed on $\sin^2\!\left(2\theta_{\mu e}\right) \equiv \sin^2\!\left(\theta_{24}\right) \sin^2\!\left(2\theta_{14}\right)$.
This is the observable to which the LSND and MiniBooNE experiments were sensitive and reported results in tension with the conventional three-neutrino scenario~\cite{Aguilar:2001ty,Aguilar-Arevalo:2013pmq}.
The results of the combined analysis~\cite{Carroll,Adamson:2016jku}, shown in Fig.~\ref{fig:stNu}(left) are consistent with the absence of sterile neutrinos.

Very short baseline reactor experiments are also sensitive to oscillation phenomena involving sterile neutrinos.  
Since the neutrino spectrum from reactors is not well understood, it is important to be able to obtain results at different baselines, ideally with the same detector.
This is one key concept behind the design of the DANSS experiment, the preliminary first results from which are not consistent with the currently favoured parameters~\cite{Danilov}.
The sensitivities of several very short baseline reactor neutrino experiments are compared in Fig.~\ref{fig:stNu}(right)~\cite{Giunti,Gariazzo:2017fdh}.

\subsection{Ultra-precise tests}

The studies of sectors of the SM discussed above are complemented by a range of ultra-precise tests of specific topics.
These include the spectroscopy of antiprotonic helium, which provides a measurement of the antiproton to electron mass ratio to a precision of $< 10^{-9}$ and allows tests of \CPT\ symmetry~\cite{Hori,Hori:2016ulw}.
These results can be interpreted as providing constraints on a fifth force, as can precision tests of gravity at short distances~\cite{Moore,Jaffe:2016fsh}.

\section{Searches for physics beyond the Standard Model}

\subsection{High \pt\ signatures}
\label{sec:highpT}

The SM is known to be an effective theory, valid only at energies below some cut-off scale. 
It is not known with certainty at what scale physics beyond the SM should emerge, but arguments based on ``naturalness'' suggest that the \tev\ scale is likely (see, for example, Ref.~\cite{Dine:2015xga}).
The use of the phrase ``natural'' in this technical sense can cause confusion.
The recent ``paella-gate'' controversy provides a tenuous analogy: one may na{\" i}vely think that the delicious Valencian dish paella may be improved by the addition of the tasty Spanish sausage chorizo, but this is incorrect -- it is {\it unnatural} to do so.
Similarly, there are only certain ways to augment the SM that a purist may consider acceptable.  

\begin{figure}[!htb]
  \centering
  \includegraphics[width=0.44\linewidth]{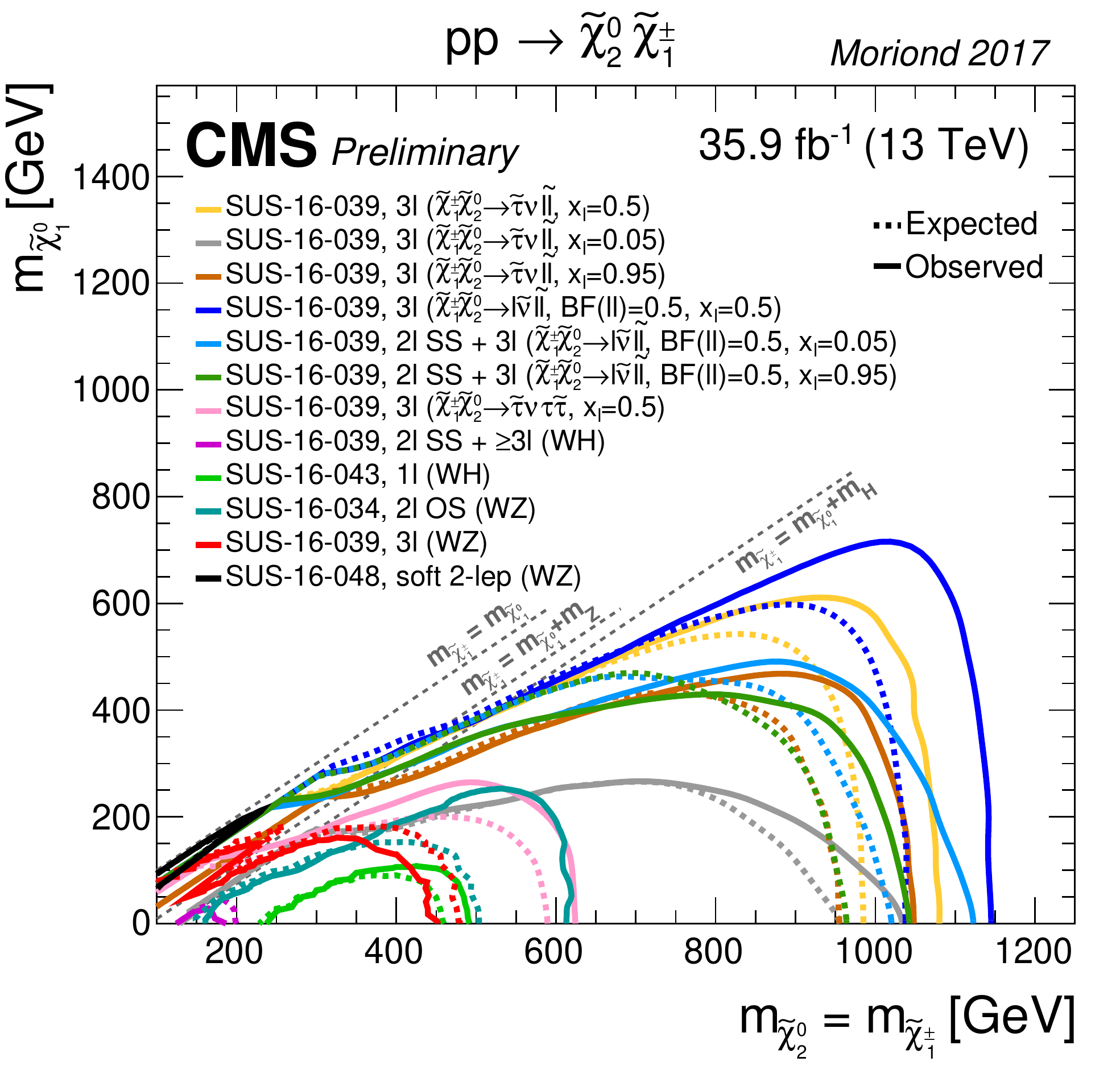}
  \includegraphics[width=0.46\linewidth]{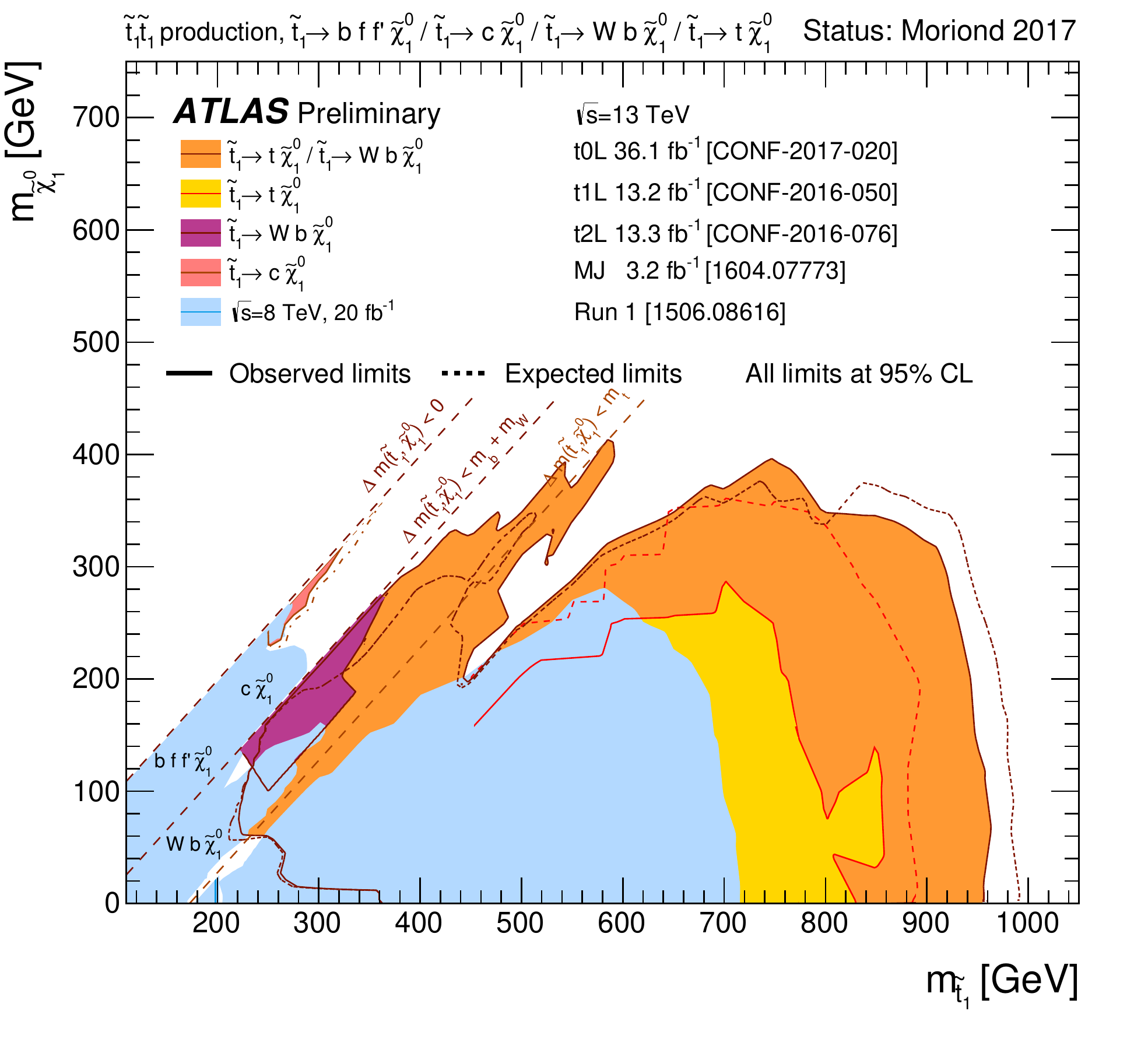}
  \caption[]{
    Limits on (left) chargino-neutralino pair production from CMS~\cite{CMS-SUSY-plots} and (right) stop pair production from ATLAS~\cite{ATLAS-SUSY-plots}.
    
    }
  \label{fig:SUSY}
\end{figure}

Supersymmetry is a good example of a ``natural'' extension to the SM, which it is therefore well motivated to search for.  
The ATLAS and CMS experiments have performed an astonishing amount of work to carry out these searches, analysing their $13 \tev$ data samples within short turn-around times from the completion of data taking~\cite{Kuwertz,Marionneau,Petridis}.  
It seems unfair to those who have invested such time and effort into this work to summarise it only with exclusion limit plots such as those shown in Fig.~\ref{fig:SUSY}; unfortunately, in the continuing absence of any evidence of a signal, it is also inevitable.

\begin{figure}[!htb]
  \centering
  \includegraphics[width=0.46\linewidth]{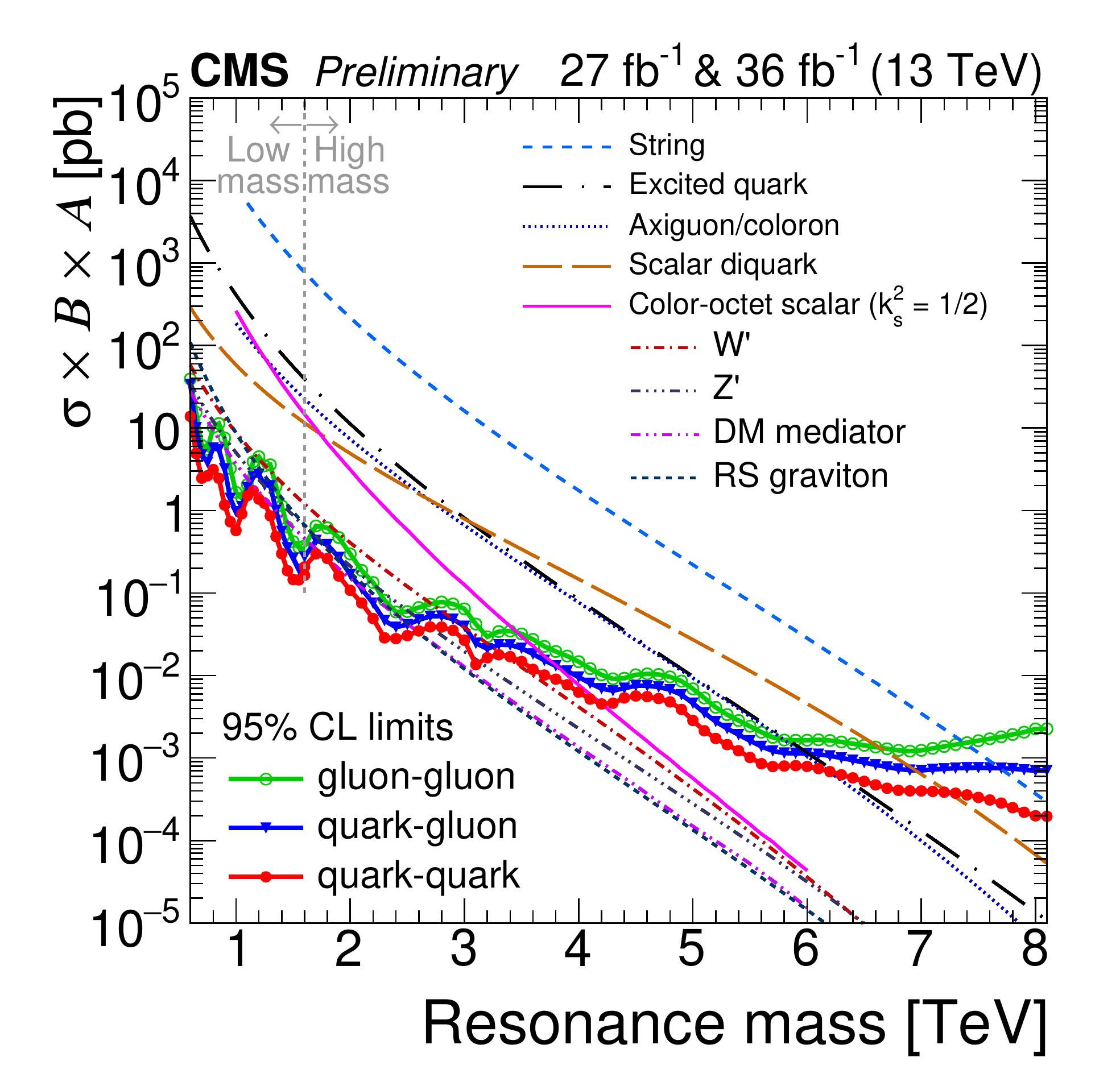}
  \includegraphics[width=0.44\linewidth]{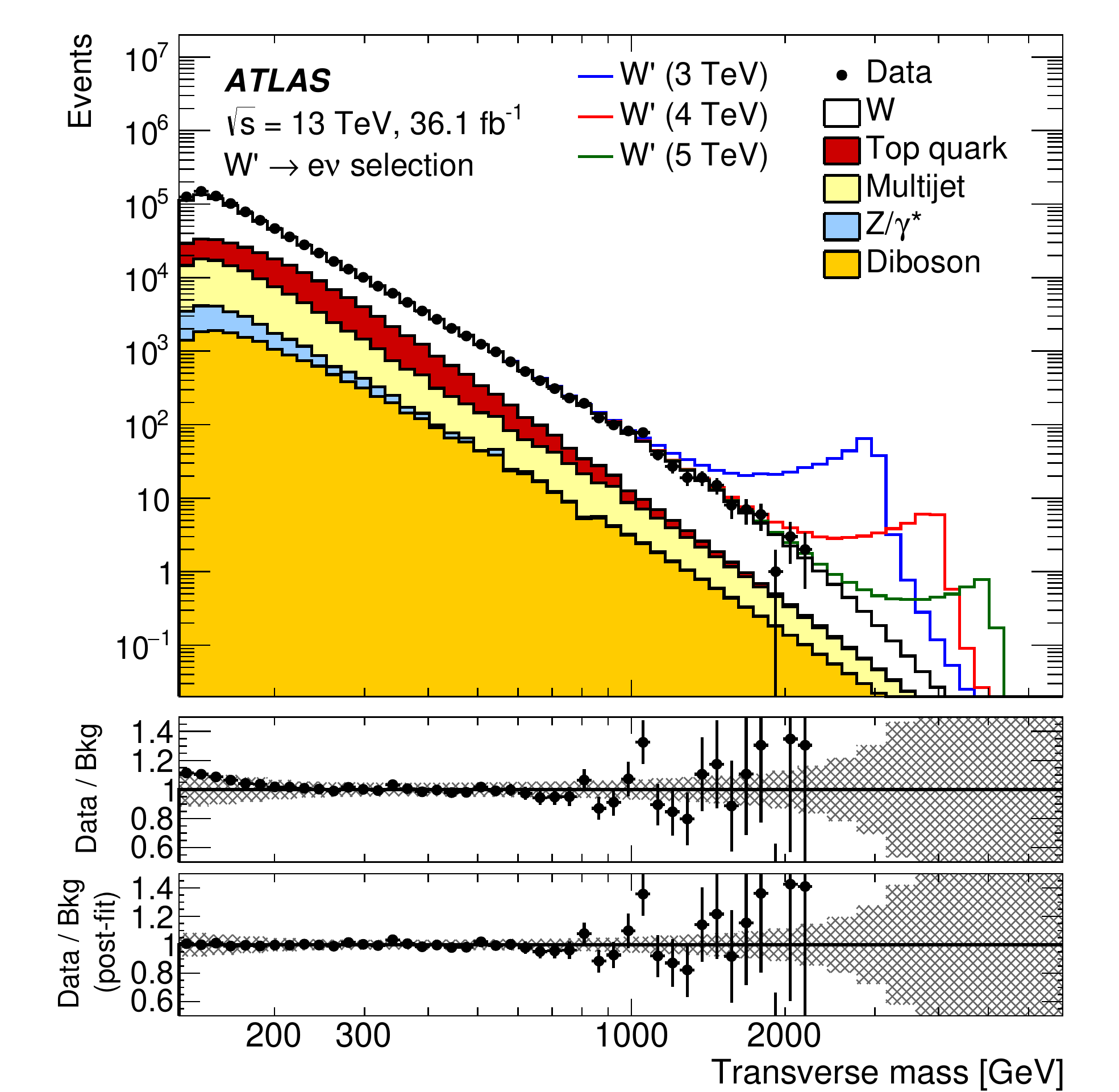}
  \caption[]{
    (Left) Upper limits at 95\% confidence level on the product of the cross-section, branching fraction, and acceptance for dijet resonances interpreted as (red) quark-quark, (blue) quark-gluon, and (green) gluon-gluon type.
    The predicted cross-sections in various models are also shown.
    (Right) Transverse mass distribution for $W^\prime \to e\nu$ candidates in ATLAS data compared to the expected backgrounds~\cite{Aaboud:2017efa}. 
    Expected signal distributions for different $W^\prime$ masses are also shown.
    }
  \label{fig:BSM}
\end{figure}

Although the searches mentioned above are characterised as being ``SUSY searches'', they may also be sensitive to other models.
Similarly, generic signatures of physics beyond the SM such as invariant mass peaks in dijet, dilepton or diboson distributions are of great interest.
The dijet final state suffers from copious background~\cite{Gao}, which has been handled in ATLAS with a sliding window mass fit~\cite{Aaboud:2017yvp} and in CMS with a novel data scouting approach for the low mass candidates~\cite{CMS-PAS-EXO-16-056}, illustrated in Fig.~\ref{fig:BSM}(left).
Among the dilepton final states, new results on $W^\prime \to e\nu$ -- shown in Fig.~\ref{fig:BSM}(right) -- and $\mu\nu$ allow limits at the level of 4--$5 \tev$ to be set~\cite{Radogna,ATLAS-CONF-2017-016,Aaboud:2017efa,Khachatryan:2016jww}.
In the diboson channels, where jet grooming techniques are becoming increasingly important to handle effects of pile-up, a small excess in the ATLAS $V\!H$ data at $\sim 3 \tev$ is not seen in the CMS data, with the excluded region reaching up to around $2.5\tev$~\cite{Li,ATLAS-CONF-2017-018,CMS-PAS-B2G-17-002}.
Many other signatures have also been investigated~\cite{TalHod}.

Since no clear sign of physics beyond the SM has emerged at the energy frontier with the $13 \tev$ LHC data, one may be tempted to conclude that the new physics energy scale is out of reach of the LHC.
However, this would be premature.
Significant increases in sensitivity anticipated with larger data samples~\cite{Shchutska,Genest}.
For example, the lower limits on the stop mass which is currently at about $1.1 \tev$ can be pushed up to $1.5 \tev$ with $300 \invfb$ and around $2 \tev$ with the full HL-LHC sample of $3000 \invfb$~\cite{Shchutska}.
However, increasingly strong limits also provide motivation to develop and pursue different ideas.
Searches for more exotic signatures, such as lepton flavour violation, can be sensitive to weaker couplings~\cite{Radogna}.

The concept of new states that are weakly coupled to the SM motivates searches for long-lived particles.
Here, ``long-lived'' may mean that these particles are effectively stable in the detectors, in which case they may be detectable through their $dE/dx$ distribution~\cite{Spiezia}.
Or it may mean that they decay inside the detector material, giving signatures of kinked or even disappearing tracks. 
The ATLAS inner B layer (IBL) that was installed during the LHC long shutdown enhances sensitivity to such models~\cite{Kaji,ATLAS-CONF-2017-017}.
The geometry of the LHCb vertex locator provides unique sensitivity to particles that may be produced in decays of heavier states (whether SM such as $B$ mesons or new heavy states) and travel $\lesssim {\cal O}\left(1 \m\right)$ before decaying~\cite{Hulsbergen,LHCb-PAPER-2016-065,LHCb-PAPER-2016-052}.
The results of all such searches to date are, however, null.

\subsection{Signatures in rare decays of SM particles}

Physics beyond the SM can also be manifest in certain rare or forbidden SM processes.
Indeed, historically there have usually been signals appearing in such processes before discoveries at the energy frontier.  
The heavy quark sector is ideal for searches for deviations from SM predictions, since: (i) the SM has a highly distinctive flavour structure that may not be replicated in extended models; (ii) for certain processes the SM predictions have low theoretical uncertainties.

\begin{figure}[!htb]
  \centering
  \includegraphics[width=0.51\linewidth]{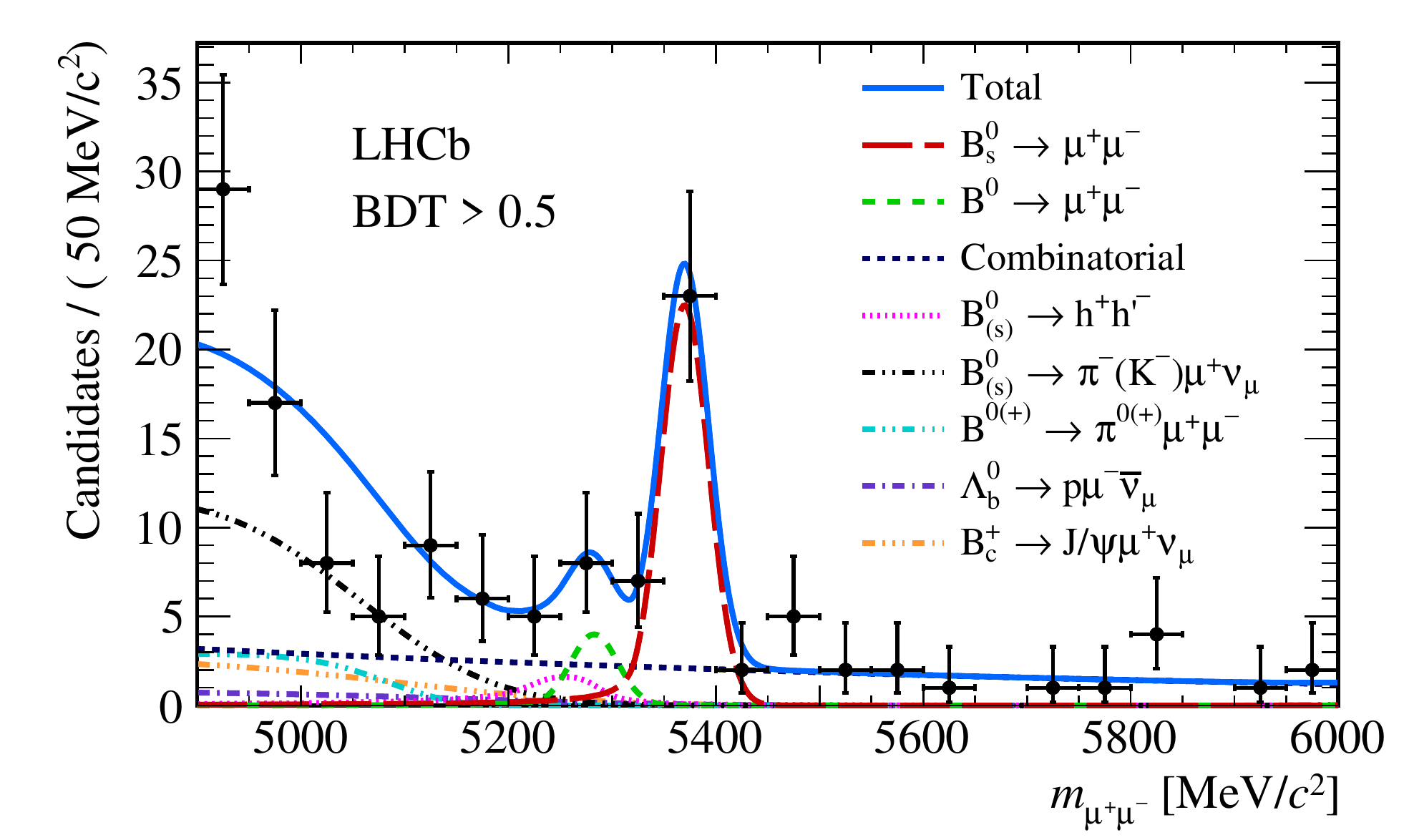}
  \includegraphics[width=0.475\linewidth]{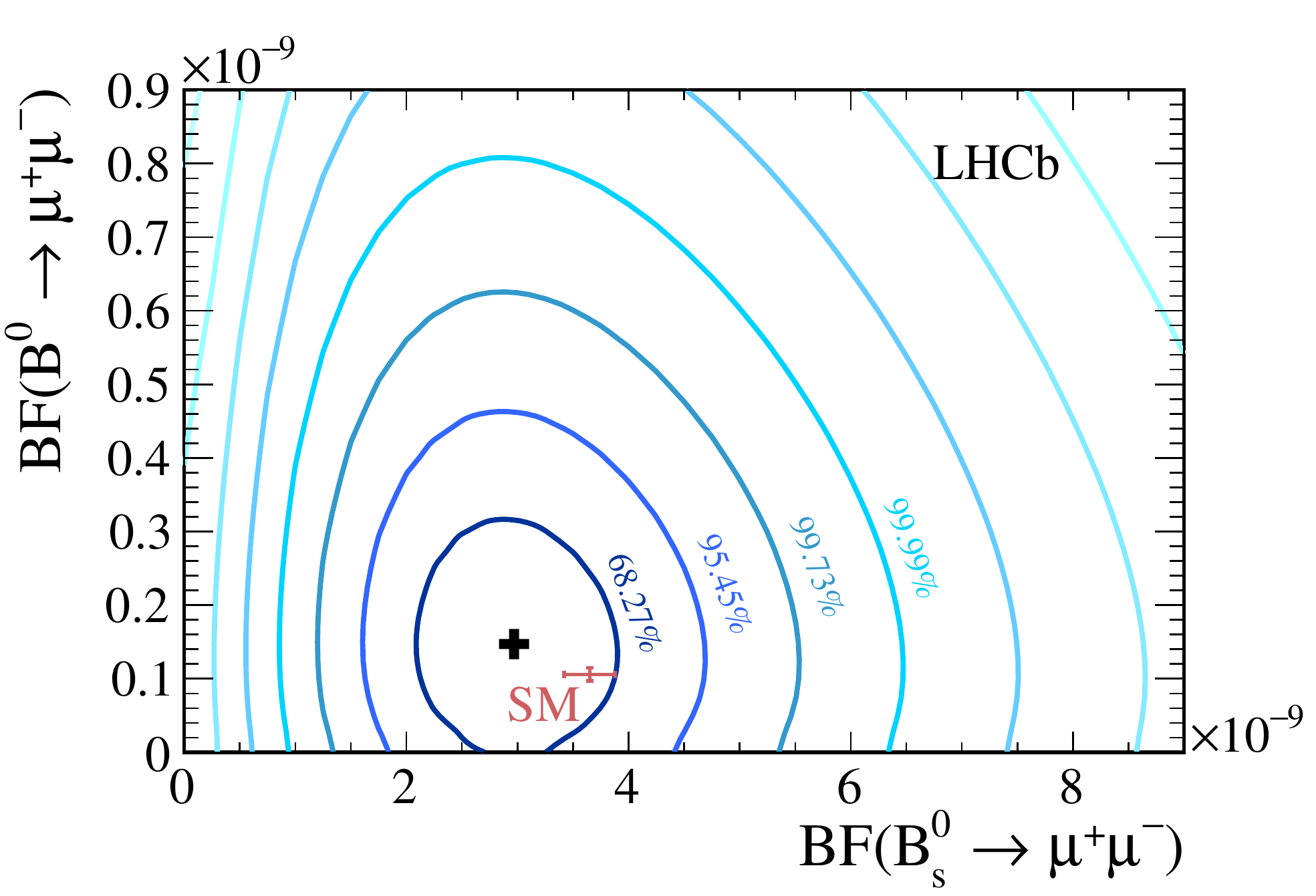}
  \caption[]{
    (Left) Dimon invariant mass distribution for \Bdors\ candidates, and (right) confidence level contours in the plane of the \Bd\ and \Bs\ dimuon branching fractions, from LHCb~\cite{LHCb-PAPER-2017-001}.
}
  \label{fig:Bmumu}
\end{figure}

Among the golden modes for such searches are the dilepton decays of $B$ mesons.  
In the SM, these are predicted to have small branching fractions due to the ``helicity suppression'' caused by the small lepton masses (compared to the $B$ mass) and the $V-A$ structure of the weak interaction.
The rates for neutral $B$ meson decays are further suppressed by the absence of tree-level flavour-changing neutral-currents in the SM.
New results from LHCb, including Run~II data, on $\Bdors \to \mumu$~\cite{Tolk,LHCb-PAPER-2017-001} give the first single experiment observation of the \Bs\ mode, as shown in Fig.~\ref{fig:Bmumu}.
(This follows a previous combined analysis of LHCb and CMS Run~I data~\cite{LHCb-PAPER-2014-049} and an independent result from ATLAS with slightly less sensitivity~\cite{Aaboud:2016ire}).
The results of all measurements are consistent, but the significance of the \Bd\ decay, which is further suppressed compared to its \Bs\ counterpart by the ratio of CKM matrix elements $\left| V_{td}/V_{ts}\right|^2$, has reduced from its previous value just touching $3\sigma$. 
LHCb has also presented the first direct limits on $\Bs \to \taup\taum$~\cite{LHCb-PAPER-2017-003}.
These results put strong constraints on models that extend the SM with new scalar or pseudoscalar interactions.  

As regards the leptonic decays of charged heavy flavour mesons, results from Belle on the $\Bp \to \taup\nu_{\tau}$ decay~\cite{Hirose,Kronenbitter:2015kls} have a signal significance of $4\sigma$, with the measured branching fraction consistent with the SM.
Measurements from BESIII of the $\Dp$ and $\Dsp$ decays to $\mup\nu_{\mu}$ and $\taup\nu_{\tau}$~\cite{Weidenkaff,Ablikim:2016duz}, including first evidence for the $\Dp \to \taup\nu_{\tau}$ mode, are in good agreement with the SM predictions that rely on lattice QCD calculations.

The pattern of branching fractions predicted by the SM for BEH boson decays is also highly distinctive, and therefore rare BEH decays are also of much interest to test the SM.  
An example is a search by CMS for $H \to a_1a_1$, where $a_1$ is a new (beyond SM) light boson that decays (in this case) into a dimuon pair~\cite{MartinezOutschoorn,CMS-PAS-HIG-16-035}.
No significant signal is seen.
Another good example is the lepton flavour violating decay mode $H \to \tau^\pm\mu^\mp$, where a slight excess had been seen by CMS in Run~I data~\cite{Khachatryan:2015kon}, which results from ATLAS~\cite{Aad:2016blu} and from a subset of CMS Run~II data~\cite{CMS-PAS-HIG-16-005} are not sensitive enough to exclude.\footnote{
  Since Moriond, an updated analysis with more CMS Run~II data~\cite{CMS-PAS-HIG-17-001} results in significantly stronger limits on the $H \to \tau^\pm\mu^\mp$ branching fraction.
}

\subsection{Dark matter searches}

There are several observational facts about our Universe that are not explained by the SM.
These range from things which have been known essentially forever (gravity exists), to things that have become known relatively recently following sophisticated obervations (there is some kind of dark energy fuelling the accelerating expansion of the Universe). 
Somewhere in between is the knowledge that there must be dark matter that explains, {\it inter alia}, the observed rotations of galaxies.
Despite the exciting discovery that ``Dark Matter'' is a Scottish spiced rum,\footnote{
  This perhaps went undetected for a long time since the idea of rum from Scotland is far outside the Standard Model of distillery.
}
there remains no certain understanding of whether the solution to this mystery has implications for particle physics.
Nonetheless, there are strong arguments that make scenarios such as the WIMP hypothesis very attractive, motivating increasingly precise searches.

\begin{figure}[!htb]
  \centering
  \includegraphics[width=0.45\linewidth]{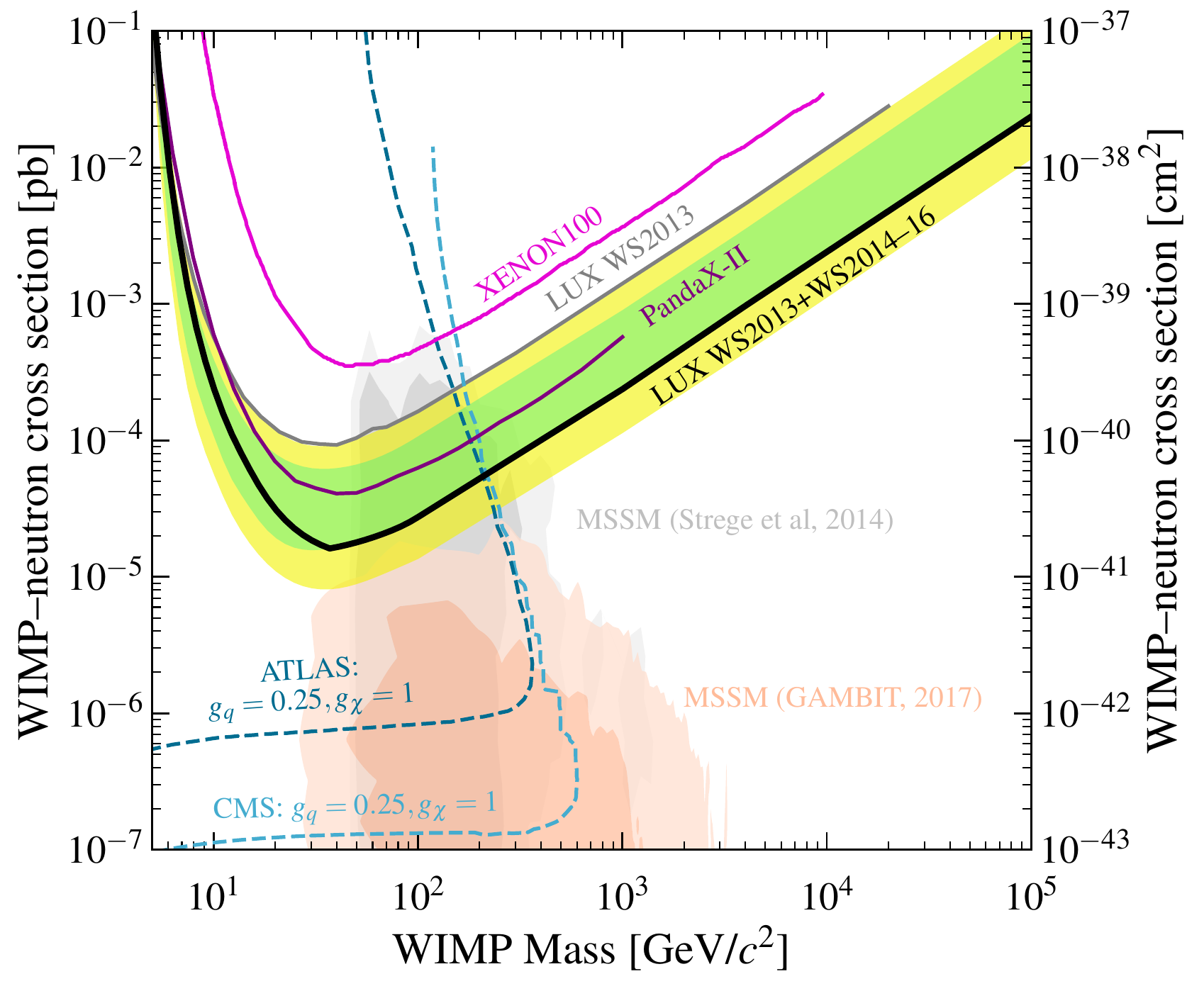}
  \hspace{0.05\linewidth}
  \includegraphics[width=0.45\linewidth]{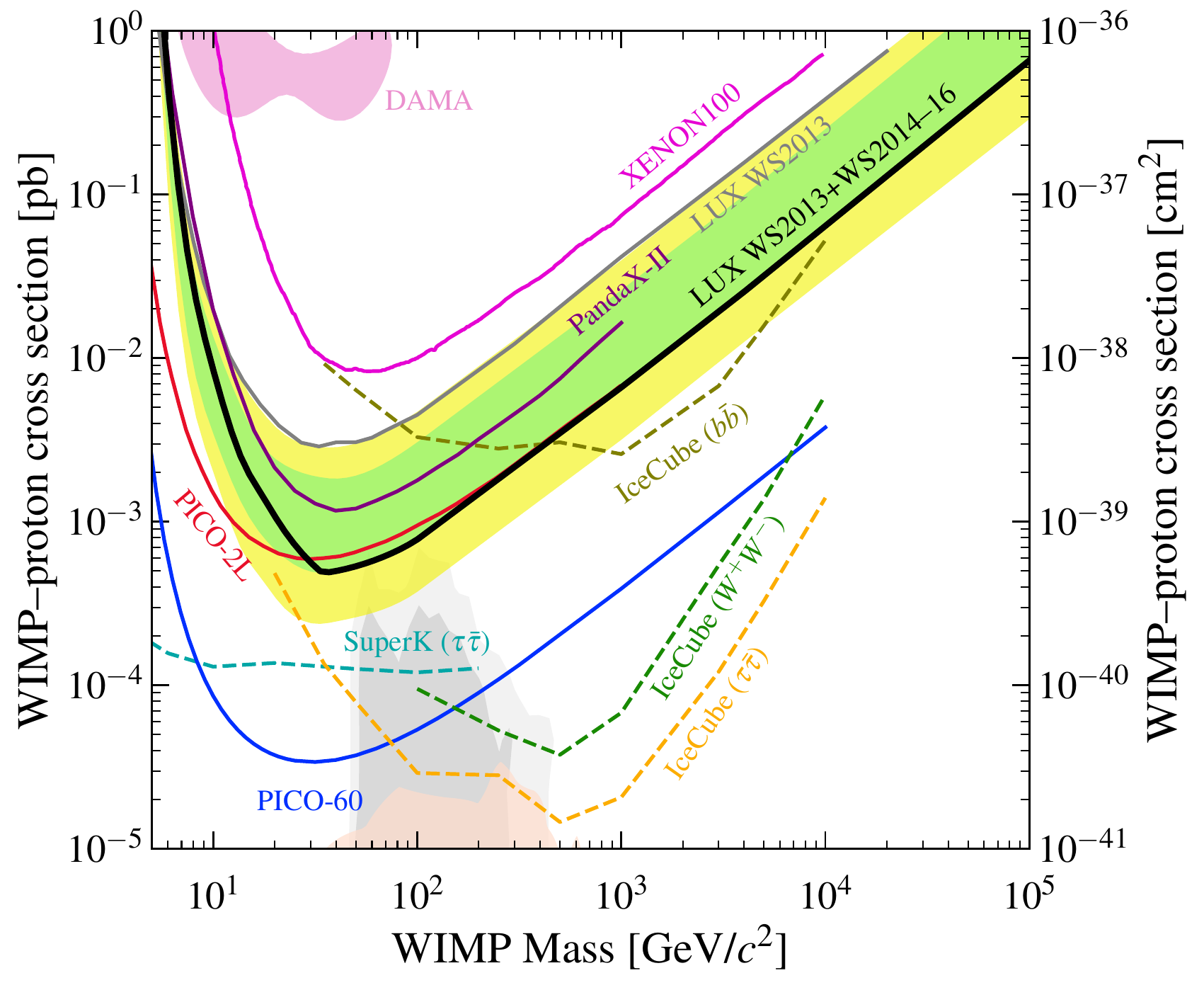}
  \caption[]{
    Upper limits on the (left) WIMP-neutron and (right) WIMP-proton spin-dependent cross-sections, at 90\% confidence level, from LUX (dark line)~\cite{Akerib:2017kat}.  
    The 68\% and 95\% ranges for the expected sensitivity are shown in green and yellow, together with results from previous liquid Xenon (PandaX-II and XENON100) and other experiments.                           
    In the WIMP-neutron case, model-dependent LHC search results are shown by dashed lines.
    The favoured parameter space in each of two MSSM analyses is shown in peach or gray.
}
  \label{fig:DM}
\end{figure}

Searches for WIMPS proceed in several complementary ways.
In case such particles are produced in high energy $pp$ collisions, they are expected to leave signatures of missing energy in the ATLAS and CMS detectors.
The results of many of the searches discussed in Sec.~\ref{sec:highpT}, or modified versions of those searches, can be interpreted as setting limits on the WIMP parameter space~\cite{Madsen,AngeloGerosa}.
This is often referred to as ``indirect detection'' in constrast to the ``direct detection'' experiments looking for signals of dark matter particles scattering of the material in their detectors.  
An example of the latter is the large scale liquid Xenon LUX detector, which has reported new spin-dependent limits, as shown in Fig.~\ref{fig:DM}~\cite{Masbou,Akerib:2017kat}.

For low mass WIMPS, different detection techniques have better sensitivity.
In the PICO60 experiment, dark matter is searched for by looking and listening (with visual and audio detectors) for bubbles that would be caused by WIMP interactions in superheated freon.\footnote{
  Since smelling and tasting the freon is unadvisable, this seems a good choice of senses to use.
}
The obtained limits~\cite{Giroux,Amole:2017dex} can be seen in Fig.~\ref{fig:DM}(right).
At even lower WIMP mass, CCDs can be used to detect ionisation caused by nuclear recoil following coherent elastic scattering of the dark matter particle.
The DAMIC experiment has obtained results using this technique~\cite{Gaior,Aguilar-Arevalo:2016ndq}, and has excellent prospects for improvement in sensitivity with larger scale detectors -- as indeed do the other approaches for WIMP searches discussed here.

Gamma ray observatories can also be used to search for signatures of dark matter annihilation.  
There are several approaches, including studies of the galactic centre, of dwarf galaxies, of galaxy clusters and of substructures in the galactic halo.
Studies of dwarf galaxies are particularly interesting as these are expected to be the most dark matter dominated systems in the Universe.
There is, however, no clear signal for dark matter annihilation in results from HESS and Fermi~\cite{Viana,HESS2011,Fermi-LAT:2016uux}.

The range of detection approaches available means that there are good prospects to detect WIMPs in the near future, if they have masses in the \gev\ to \tev\ range and scattering cross-sections not too far below the current bounds (as predicted in many models).  
The scaling of sensitivity achieved with background-free detectors cannot continue indefinitely due to the so-called ``neutrino floor'' due to coherent scattering off the detector material of solar, atmospheric or supernova neutrinos, but this is not yet a limitation.
However, since much of the preferred phase-space for certain models has been ruled out there is also good justification to consider alternative explanations of dark matter, including axions and other scenarios~\cite{Gavela}.
One interesting class of models contain a dark sector, which may mix with the SM through a dark photon.
The BaBar experiment has performed a range of searches for dark photons, the most recent of which targets their invisible decay, and hence is particularly sensitive in case the coupling to the SM, $\epsilon$, is small.
Limits on $\epsilon$ at the level of $10^{-3}$ up to dark photon masses of around $10 \gev$ are obtained~\cite{Rohrken,Lees:2017lec}.

\subsection{Understanding the origin of neutrino mass}

\begin{figure}[!htb]
  \centering
  \includegraphics[width=0.35\linewidth]{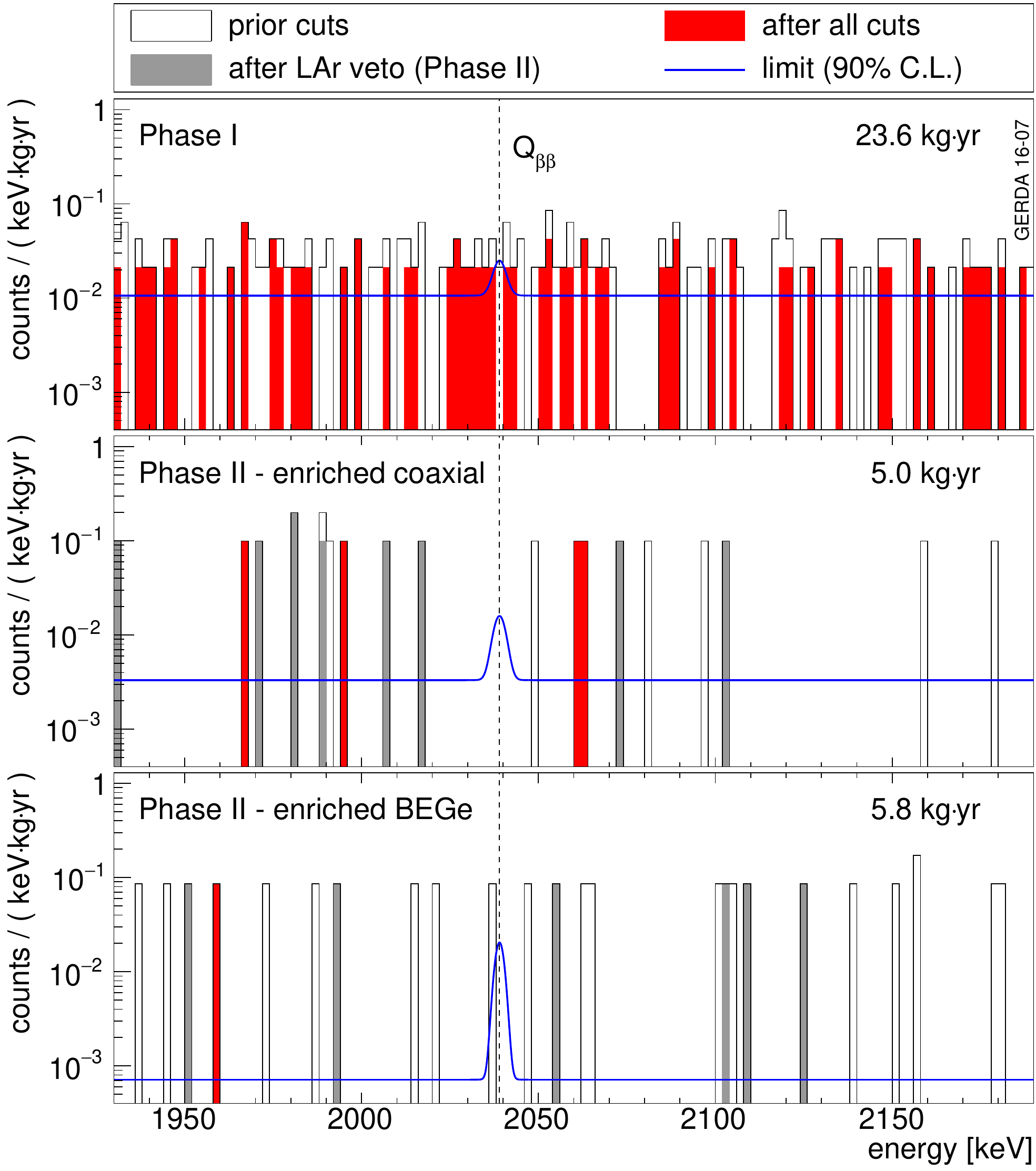}
  \includegraphics[width=0.55\linewidth]{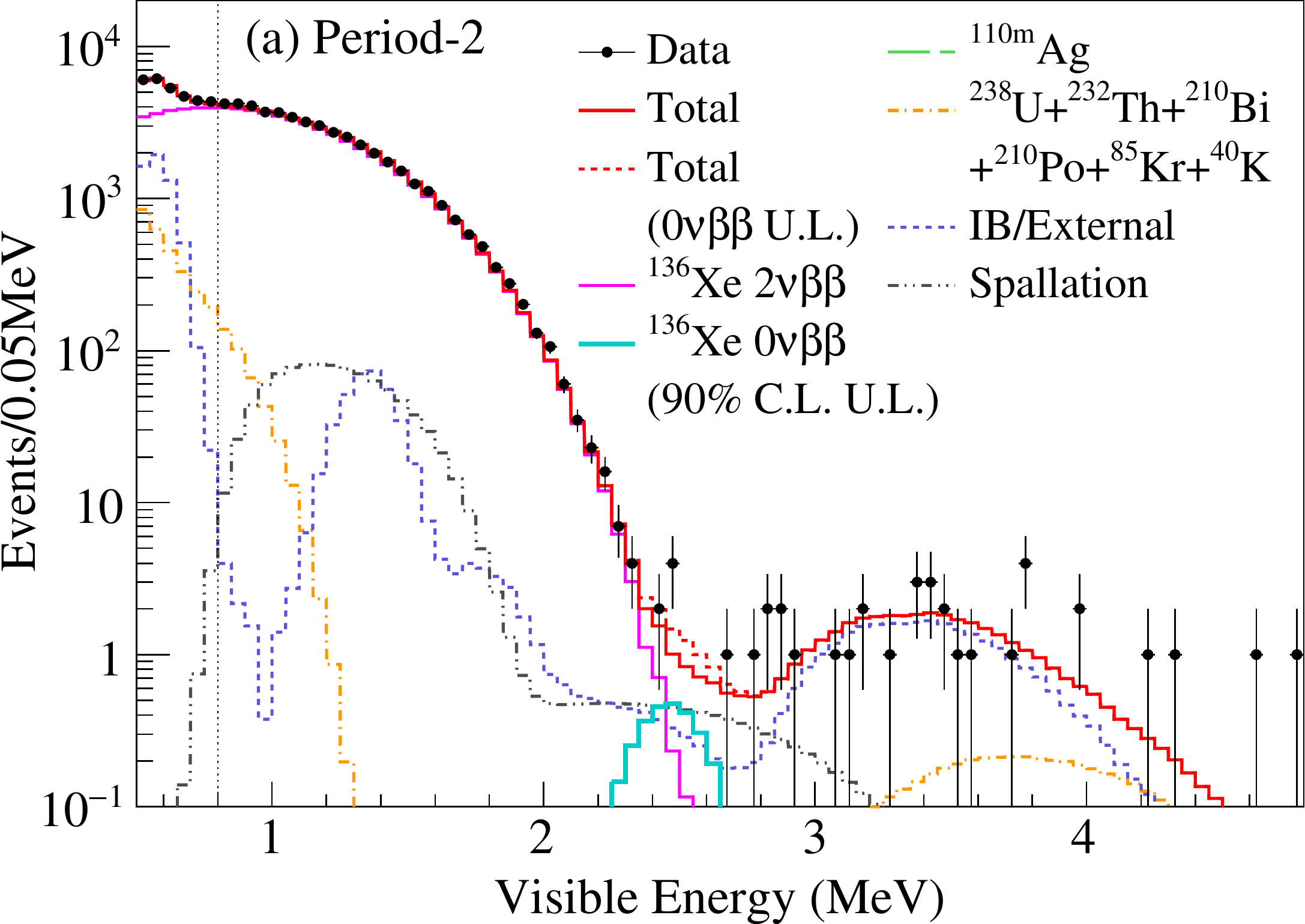}
  \caption[]{
    (Left) Data from GERDA Phase~I and ~II showing the data to be almost background free in Phase~II after all cuts (red histogram)~\cite{Agostini:2017iyd}.
    (Right) Energy spectrum of selected $\beta\beta$ candidates in KamLAND-Zen Period 2 together with the best fit curve including components from different backgrounds as described in the legend~\cite{KamLAND-Zen:2016pfg}.
    In both cases a signal peak is imposed on the background fit corresponding to the 90\% confidence level lower limit on $T^{0\nu}_{1/2}$.
}
  \label{fig:nuMass}
\end{figure}

As mentioned previously, neutrinos are massless in the SM and therefore it is a pressing question as to how their mass should be explained.
This is arguably the most fundamental question in particle physics, since it may involve extending the SM by introducing a new type of field: a Majorana fermion.
The most sensitive way to test the nature of neutrino mass is through neutrinoless double beta decay, which is being pursued by a number of highly sophisticated experiments worldwide.
The GERDA experiment has achieved a preliminary 90\% confidence level lower limit of the neutrinoless double beta decay half-life $T^{0\nu}_{1/2} > 5.3 \times 10^{25} \ {\rm yr}$ in $^{76}{\rm Ge}$~\cite{Wagner,Agostini:2017iyd}, as shown in Fig.~\ref{fig:nuMass}(left).
The KamLAND-Zen experiment has achieved (after dealing with $^{100{\rm m}}{\rm Ag}$ background due to nuclear fallout from the Fukushima catastrophe) a limit of $T^{0\nu}_{1/2} > 1.1 \times 10^{26} \ {\rm yr}$ in $^{136}{\rm Xe}$~\cite{Decowski,KamLAND-Zen:2016pfg}, as shown in Fig.~\ref{fig:nuMass}(right).
Translating these results into constraints on the effective neutrino mass requires knowledge of nuclear matrix elements and is therefore somewhat model dependent, but the limits are starting to approach the region where signals are expected in the case of inverted neutrino mass hierarchy.  
Significant improvement is expected with the inclusion in the analysis of more data in the current experimental configuration, with more data that will be recorded after detector upgrades, and (longer term) with even more data that can be obtained with larger scale experiments.

\section{Anomalies (of varying significance) -- a.k.a. ``pink unicorns''}

While the overall picture is one of consistency with the SM, and no clear signal of physics beyond the SM emerging (apart from the longstanding areas of dark matter and neutrino mass), there are indications in several observables that -- optimistically -- could potentially evolve into discoveries as the precision improves.
Of course, this is always the case: with the huge number of measurements being made, it is to be expected that one or two may have significance above $3\sigma$ just due to (statistical or systematic) fluctuations.
Thus, it is essential to keep in mind that the $5\sigma$ threshold usually required to claim a discovery in particle physics is there for a reason.  

\begin{figure}[!htb]
  \centering
  \includegraphics[width=0.49\linewidth]{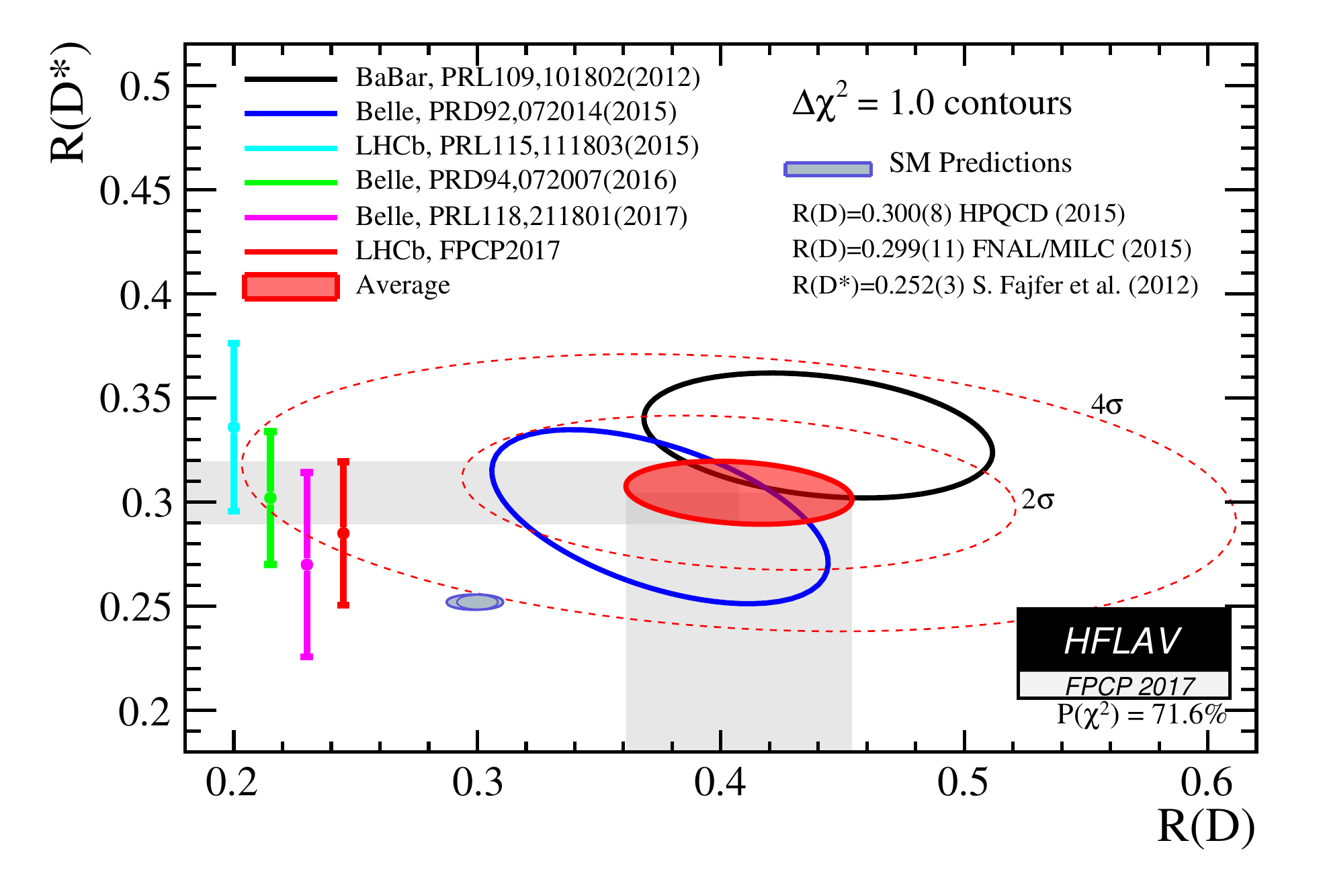}
  \includegraphics[width=0.49\linewidth]{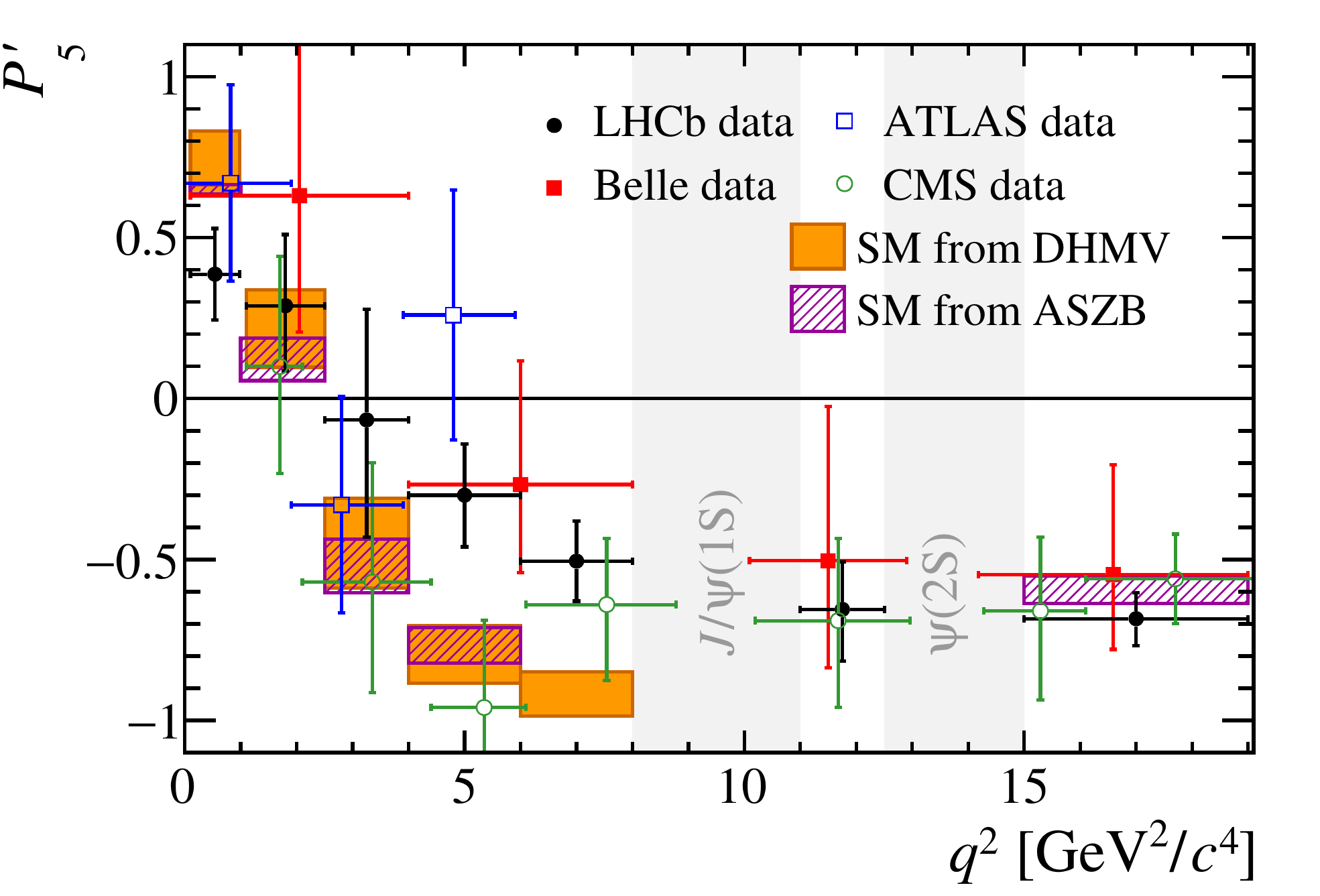}
  \caption[]{
    (Left) World average of measurements of $R(D)$ and $R(D^*)$, accounting for their correlations, compared to the SM prediction~\cite{HFLAV}.
    (Right) Measurements of the observable $P_5^\prime$ in bins of $q^2$ from ATLAS, CMS, LHCb and Belle, compared to SM predictions.
}
  \label{fig:unicorns}
\end{figure}

There are indications of violation of lepton universality in semileptonic $B$ decays.
The observables $R(D^{(*)}) \equiv {\cal B}\left(B \to D^{(*)}\tau\nu_{\tau}\right)/{\cal B}\left(B \to D^{(*)}\ell\nu_{\ell}\right)$ have been found to be larger than their SM expectations by all of BaBar, Belle and LHCb~\cite{Wormser,Hirose,Lees:2012xj,Lees:2013uzd,Huschle:2015rga,Sato:2016svk,Hirose:2016wfn,LHCb-PAPER-2015-025,LHCb-PAPER-2017-017},\footnote{
  Results from LHCb on $B \to D^{*}\tau\nu_{\tau}$ with the $\tau$ lepton reconstructed in the three-prong final state~\cite{LHCb-PAPER-2017-017}, which at Moriond were ``coming soon'', have been included in this discussion.
}
with the world average shown in Fig.~\ref{fig:unicorns}.
(In the definition of $R(D^{(*)})$, $\ell$ can be either electron of muon, although only muons are used in the LHCb measurements.)
The discrepancy between the average of the experimental measurements, which are self-consistent, and the SM prediction is at the level of $4\sigma$.
More precise measurements are clearly needed.

Also related to lepton universality violation are the similar ratios for the neutral current semileptonic decay channels: $R(K^{(*)})  \equiv {\cal B}\left(B \to K^{(*)}\mumu\right)/{\cal B}\left(B \to K^{(*)}\epem\right)$.  
Within the SM this ratio should be equal to one up to small corrections due to lepton masses and QED effects.  
LHCb has measured $R(K)$ in the $q^2 = m^2(\ell^+\ell^-)$ bin from $1$ to $6 \gevgevcccc$, finding a value $2.6\sigma$ below the SM~\cite{Bifani,LHCB-PAPER-2014-024}.
New results\footnote{
  The LHCb results on $R(K^*)$, which at Moriond were ``coming soon''~\cite{Bifani}, have been included in this discussion. 
} 
on $R(K^*)$ in the bins $0.045 < q^2 < 1.1 \gevgevcccc$ and $1.1 < q^2 < 6.0 \gevgevcccc$ are also below the SM prediction~\cite{LHCb-PAPER-2017-013}, each with significance of $2$--$2.5\sigma$.

These results complement the numerous other observables that have been studied in $B \to K^{(*)}\ell^+\ell^-$ decays, including differential branching fractions and angular observables such as the infamous $P_5^\prime$ (recent reviews can be found in Refs.~\cite{Blake:2015tda,Blake:2016olu}).
New results from ATLAS~\cite{Bevan,ATLAS-CONF-2017-023} and CMS~\cite{Dinardo,CMS-PAS-BPH-15-008} complement previous measurements from LHCb~\cite{LHCb-PAPER-2015-051} and Belle~\cite{Wehle:2016yoi} and add to the picture in Fig.~\ref{fig:unicorns}(right).
The averages of the data points in the bins $4 < q^2 < 6 \gevgevcccc$ and $6 < q^2 < 8 \gevgevcccc$ are noticeably above the SM central values, but the significance depends on the uncertainty assigned to the prediction.
This is hard to quantify due to possible effects due to charm loops (which also explain why some experiments and theory groups do not provide data in the bin just below the $\jpsi$ region).
However, it is important to note that data-driven progress can be made to address this issue, for example by studying interference effects with the charmonium resonances~\cite{LHCb-PAPER-2016-045}.

There are several other discrepancies between data and SM predictions that are worth recapping.
The anomalous muon magnetic moment, $(g-2)_\mu$, is measured to differ by $3.6\sigma$ from the SM value~\cite{Bennett:2006fi}; this has held up under numerous reappraisals of the theoretical uncertainty.
A new experiment will reduce the experimental uncertainty by a factor of about 2, which together with improvement in the theory should be able to clarify the situation.
The DAMA/LIBRA experiment observed an annual modulation of a possible signal for dark matter interactions~\cite{Bernabei:2013xsa}.
Results from the XENON100 experiment however significantly disfavour interpretations where this signal is due to WIMP scattering~\cite{Masbou,Aprile:2017yea}.
A ``proton radius puzzle'' has emerged from precise new results of the spectroscopy of muonic hydrogen~\cite{Antognini,Antognini:1900ns} compared to previous results from $e^-p$ scattering and $H$ spectroscopy; however further work is needed to rule out experimental issues causing the discrepancy.
Finally, new lattice QCD predictions of the parameter of direct \CP\ violation in the neutral kaon system, $\epsilon^\prime/\epsilon$ give a value about $2.8\sigma$ below the measured value~\cite{Soni,Nierste}.
Unfortunately this is one area where no new experiments are currently planned, so it may be difficult to confirm if there is a non-SM contribution to this observable, even if the theory predictions improve further (as is also much needed).
However, there are good prospects for measurements of the rare kaon decays $K^+ \to \pip\nu\bar{\nu}$ at NA62 and $K_L \to \piz\nu\bar{\nu}$ at K0T0~\cite{Lurkin}.
Results from these experiments could potentially be among the highlights of the $53^{\rm rd}$ Rencontres de Moriond in March 2018.

\section{Summary}

One striking feature of all the different topics discussed during the Rencontres de Moriond is the excellent prospects for improvements in sensitivity.
The LHC is performing superbly, and upgrades of the accelerator and detectors are planned.  
The SuperKEKB accelerator will soon start providing physics quality $\epem$ collisions to the Belle~II detector. 
Next generation experiments are being developed for both long baseline and reactor neutrino oscillations, as well as to investigate the mystery of neutrino mass.
Dark matter experiments are likewise preparing to make the next jump in scale, and there are numerous new facilities being constructed to investigate the cosmic frontier.  
Thus, it is clear that the Moriond series will remain a forum for lively discussions of exciting new results for many years to come.

Will the future bring us another dramatic new discovery, comparable to that of BEH boson five years ago?
The answer may well depend on interpretations of the word ``comparable''.  
The BEH discovery was striking for many reasons, but not least because of its signature as a clear invariant mass peak, on top of a smoothly varying background.
This is an ideal discovery scenario, not only for explaining to the general public, but for interpretation and acceptance within the particle physics community.
It allowed the then Director General of CERN to make the simple declaration: ``I think we have it.''
But even if nature is kind enough to leave signatures of physics beyond the SM within our grasp, and indeed the prospects for discoveries in the coming years remain excellent, it may not be so generous to make them of exactly the type that we most hope for.  
Perhaps we will spend months or years not quite sure if we ``have it'' or not.
So, bearing in mind that science is the new rock and roll~\cite{Cox}, we should perhaps take inspiration from the Rolling Stones: ``You can't always get what you want ... but if you try ... you might find ... you get what you need.''
We must keep trying.

\section*{Acknowledgments}

I would like to express my deepest gratitude to the organisers for bestowing upon me the great honour and challenge of presenting this summary.
I would also like to thank all Moriond participants for their stimulating talks, and for interesting discussions and explanations which helped me to understand much of the material presented.
Any errors in the summary are, of course, my own.  
This work is supported by 
the Science and Technology Facilities Council (United Kingdom).

\section*{References}
\setboolean{inbibliography}{true}
\bibliographystyle{LHCb}
\bibliography{references,moriond,LHCb-main,LHCb-PAPER,LHCb-CONF}

\end{document}

%% file: lhcb-symbols-def.tex
\usepackage{xspace} 
\usepackage{upgreek}

\def\MagUp {\mbox{\em Mag\kern -0.05em Up}\xspace}

\ifthenelse{\boolean{uprightparticles}}%
{

 \def\Pmu         {\ensuremath{\upmu}\xspace}

 \def\Ppi         {\ensuremath{\uppi}\xspace}

 \def\Ptau        {\ensuremath{\uptau}\xspace}

 \def\Ppsi        {\ensuremath{\uppsi}\xspace}

 \def\PDelta      {\ensuremath{\Delta}\xspace}                 
 \def\PXi      {\ensuremath{\Xi}\xspace}                 
 \def\PLambda      {\ensuremath{\Lambda}\xspace}                 
 \def\PSigma      {\ensuremath{\Sigma}\xspace}                 
 \def\POmega      {\ensuremath{\Omega}\xspace}                 
 \def\PUpsilon      {\ensuremath{\Upsilon}\xspace}                 
 
 \def\PB      {\ensuremath{\mathrm{B}}\xspace}                 
                  
 \def\PD      {\ensuremath{\mathrm{D}}\xspace}

 \def\PJ      {\ensuremath{\mathrm{J}}\xspace}                 
 \def\PK      {\ensuremath{\mathrm{K}}\xspace}

 \def\Pe      {\ensuremath{\mathrm{e}}\xspace}

 \def\Pi      {\ensuremath{\mathrm{i}}\xspace}

 \def\Ps      {\ensuremath{\mathrm{s}}\xspace}

}
{

 \def\Pmu         {\ensuremath{\mu}\xspace}

 \def\Ppi         {\ensuremath{\pi}\xspace}

 \def\Ptau        {\ensuremath{\tau}\xspace}

 \def\Ppsi        {\ensuremath{\psi}\xspace}                 
                  
 \mathchardef\PDelta="7101
 \mathchardef\PXi="7104
 \mathchardef\PLambda="7103
 \mathchardef\PSigma="7106
 \mathchardef\POmega="710A
 \mathchardef\PUpsilon="7107
                  
 \def\PB      {\ensuremath{B}\xspace}                 
                  
 \def\PD      {\ensuremath{D}\xspace}

 \def\PJ      {\ensuremath{J}\xspace}                 
 \def\PK      {\ensuremath{K}\xspace}

 \def\Pe      {\ensuremath{e}\xspace}

 \def\Pi      {\ensuremath{i}\xspace}

 \def\Ps      {\ensuremath{s}\xspace}

}

\makeatletter
\ifcase \@ptsize \relax
  \newcommand{\miniscule}{\@setfontsize\miniscule{4}{5}}
\or
  \newcommand{\miniscule}{\@setfontsize\miniscule{5}{6}}
\or
  \newcommand{\miniscule}{\@setfontsize\miniscule{5}{6}}
\fi
\makeatother

\DeclareRobustCommand{\optbar}[1]{\shortstack{{\miniscule (\rule[.5ex]{1.25em}{.18mm})}
  \\ [-.7ex] $#1$}}

\def\epem       {{\ensuremath{\Pe^+\Pe^-}}\xspace}

\def\mup        {{\ensuremath{\Pmu^+}}\xspace}
\def\mumu       {{\ensuremath{\Pmu^+\Pmu^-}}\xspace}

\def\taup       {{\ensuremath{\Ptau^+}}\xspace}
\def\taum       {{\ensuremath{\Ptau^-}}\xspace}

\def\squark    {{\ensuremath{\Ps}}\xspace}

\def\pion   {{\ensuremath{\Ppi}}\xspace}
\def\piz    {{\ensuremath{\pion^0}}\xspace}

\def\pip    {{\ensuremath{\pion^+}}\xspace}
\def\pim    {{\ensuremath{\pion^-}}\xspace}

\def\kaon    {{\ensuremath{\PK}}\xspace}
  \def\Kbar    {{\kern 0.2em\overline{\kern -0.2em \PK}{}}\xspace}

\def\KorKbar    {\kern 0.18em\optbar{\kern -0.18em K}{}\xspace}

\def\Kp      {{\ensuremath{\kaon^+}}\xspace}
\def\Km      {{\ensuremath{\kaon^-}}\xspace}

\def\KS      {{\ensuremath{\kaon^0_{\mathrm{ \scriptscriptstyle S}}}}\xspace}

  \def\Dbar    {{\kern 0.2em\overline{\kern -0.2em \PD}{}}\xspace}
\def\D       {{\ensuremath{\PD}}\xspace}

\def\DorDbar    {\kern 0.18em\optbar{\kern -0.18em D}{}\xspace}
\def\Dz      {{\ensuremath{\D^0}}\xspace}
\def\Dzb     {{\ensuremath{\Dbar{}^0}}\xspace}
\def\Dp      {{\ensuremath{\D^+}}\xspace}

\def\Dsp     {{\ensuremath{\D^+_\squark}}\xspace}

\def\B       {{\ensuremath{\PB}}\xspace}
\def\Bbar    {{\ensuremath{\kern 0.18em\overline{\kern -0.18em \PB}{}}}\xspace}

\def\BorBbar    {\kern 0.18em\optbar{\kern -0.18em B}{}\xspace}

\def\Bu      {{\ensuremath{\B^+}}\xspace}

\def\Bp      {{\ensuremath{\Bu}}\xspace}

\def\Bd      {{\ensuremath{\B^0}}\xspace}
\def\Bs      {{\ensuremath{\B^0_\squark}}\xspace}
\def\Bdors   {{\ensuremath{\B^0_{(\squark)}}}\xspace}

\def\jpsi     {{\ensuremath{{\PJ\mskip -3mu/\mskip -2mu\Ppsi\mskip 2mu}}}\xspace}

  \def\Y#1S{\ensuremath{\PUpsilon{(#1S)}}\xspace}

\def\Lbar        {{\ensuremath{\kern 0.1em\overline{\kern -0.1em\PLambda}}}\xspace}
\def\LorLbar    {\kern 0.18em\optbar{\kern -0.18em \PLambda}{}\xspace}

\def\to                 {\ensuremath{\rightarrow}\xspace}

\def\CP                {{\ensuremath{C\!P}}\xspace}
\def\CPT               {{\ensuremath{C\!PT}}\xspace}

\def\AT#1     {\ensuremath{A_{\mathrm{T}}^{#1}}\xspace}           

\def\C#1      {\ensuremath{\mathcal{C}_{#1}}\xspace}                       
\def\Cp#1     {\ensuremath{\mathcal{C}_{#1}^{'}}\xspace}                    
\def\Ceff#1   {\ensuremath{\mathcal{C}_{#1}^{\mathrm{(eff)}}}\xspace}        
\def\Cpeff#1  {\ensuremath{\mathcal{C}_{#1}^{'\mathrm{(eff)}}}\xspace}       
\def\Ope#1    {\ensuremath{\mathcal{O}_{#1}}\xspace}                       
\def\Opep#1   {\ensuremath{\mathcal{O}_{#1}^{'}}\xspace}                    

\newcommand{\tev}{\ifthenelse{\boolean{inbibliography}}{\ensuremath{~T\kern -0.05em eV}}{\ensuremath{\mathrm{\,Te\kern -0.1em V}}}\xspace}
\newcommand{\gev}{\ensuremath{\mathrm{\,Ge\kern -0.1em V}}\xspace}
\newcommand{\mev}{\ensuremath{\mathrm{\,Me\kern -0.1em V}}\xspace}
\newcommand{\kev}{\ensuremath{\mathrm{\,ke\kern -0.1em V}}\xspace}
\newcommand{\ev}{\ensuremath{\mathrm{\,e\kern -0.1em V}}\xspace}
\newcommand{\gevc}{\ensuremath{{\mathrm{\,Ge\kern -0.1em V\!/}c}}\xspace}
\newcommand{\mevc}{\ensuremath{{\mathrm{\,Me\kern -0.1em V\!/}c}}\xspace}
\newcommand{\gevcc}{\ensuremath{{\mathrm{\,Ge\kern -0.1em V\!/}c^2}}\xspace}
\newcommand{\gevgevcccc}{\ensuremath{{\mathrm{\,Ge\kern -0.1em V^2\!/}c^4}}\xspace}
\newcommand{\mevcc}{\ensuremath{{\mathrm{\,Me\kern -0.1em V\!/}c^2}}\xspace}

\def\m    {\ensuremath{\mathrm{ \,m}}\xspace}

\def\invfb   {\ensuremath{\mbox{\,fb}^{-1}}\xspace}

\newcommand{\stat}{\ensuremath{\mathrm{\,(stat)}}\xspace}
\newcommand{\syst}{\ensuremath{\mathrm{\,(syst)}}\xspace}

\def\gsim{{~\raise.15em\hbox{$>$}\kern-.85em
          \lower.35em\hbox{$\sim$}~}\xspace}
\def\lsim{{~\raise.15em\hbox{$<$}\kern-.85em
          \lower.35em\hbox{$\sim$}~}\xspace}

\def\pt         {\mbox{$p_{\mathrm{ T}}$}\xspace}

\def\tell1  {TELL1\xspace}
\def\ukl1   {UKL1\xspace}

\newcommand{\ie}{\mbox{\itshape i.e.}\xspace}

\newcommand{\vs}{\mbox{\itshape vs.}\xspace}